\documentclass[12pt,preprint]{aastex}
\usepackage{url}
\usepackage{natbib}
\usepackage{aas_macros}
\usepackage{hyperref}
\shorttitle{Realistic modeling of local dynamo processes on the Sun}

\title{Realistic modeling of local dynamo processes on the Sun}

\author{I.~N. Kitiashvili$^{1}$, A.~G. Kosovichev$^{2}$, N.~N. Mansour$^{1}$, A.~A. Wray$^{1}$}
\affil{$^1$NASA Ames Research Center, Moffett Field, Mountain View, CA 94035, USA}
\altaffiltext{1}{e-mail: irina.n.kitiashvili@nasa.gov}
\affil{$^2$New Jersey Institute of Technology, Newark, NJ 07102, USA}

\begin{document}
\begin{abstract}
Magnetic fields are usually observed in the quiet Sun as small-scale elements that cover the entire solar surface (the `salt and pepper' patterns in line-of-sight magnetograms). By using 3D radiative MHD numerical simulations we find that these fields result from a local dynamo action in the top layers of the convection zone, where extremely weak `seed' magnetic fields (e.g., from a $10^{-6}$~G) can locally grow above the mean equipartition field, to a stronger than 2000~G field localized in magnetic structures. Our results reveal that the magnetic flux is predominantly generated in regions of small-scale helical downflows. We find that the local dynamo action takes place mostly in a shallow, about 500~km deep, subsurface layer, from which the generated field is transported into the deeper layers by convective downdrafts.
We demonstrate that the observed dominance of vertical magnetic fields at the photosphere and horizontal fields above the photosphere can be explained by small-scale magnetic loops produced by the dynamo. Such small-scale loops play an important role in the structure and dynamics of the solar atmosphere and that their detection in observations is critical for understanding the local dynamo action on the Sun.
\end{abstract}

\section{Introduction}
The origin of magnetic field generation is a key problem for understanding solar variability across a wide range of scales. Modern high-resolution observations of global magnetic fields by  the Helioseismic and Magnetic Imager (HMI) on NASA's Solar Dynamics Observatory (SDO) \citep{Scherrer2012}, as well as investigations of small-scale magnetic fields in selected areas using the Solar Optical Telescope (SOT) on {\it Hinode} \citep{Tsuneta2008} and the balloon-borne SUNRISE observatory \citep{Solanki2010}, have revealed complicated dynamics of magnetic fields on the solar surface, and their tendency to self-organize into compact magnetic structures.

Traditionally, the solar dynamo problem is divided according to two time scales: the global dynamo, operating on the scale of the 22-year solar cycle and controlling the global toroidal and poloidal fields \citep[e.g.][]{Choudhuri1995,Dikpati1999,Brandenburg2012}, and the local dynamo \citep[e.g.][]{Nordlund1992,Cattaneo1999,Cattaneo2003,Vogler2007,Kitiashvili2013b,Rempel2014}, which operates on the time scales of granulation and super-granulation from a few minutes to a few hours and is believed to be responsible for the Sun's `magnetic carpet' \citep{Schrijver1998,Schrijver2002}. However, the energetic and dynamic connections between the two dynamos are not clear.

In the global dynamo, the kinetic helicity is produced by the action of the Coriolis force on convective turbulence \citep[for recent formulations of the global dynamo theory, see e.g.][]{Brandenburg2005b,Pipin2008a,Pipin2014}. However, in the local dynamo problem, without global rotation, the mean helicity integrated over a volume is zero. This has caused a debate about the existence of a `fluctuation dynamo' driven by non-helical forced turbulence. Recent numerical simulations convincingly demonstrate the existence of such a dynamo \citep{Cattaneo1999,Schekochihin2004,Iskakov2007}. The exact physical mechanism of the fluctuating dynamo is not yet understood, and current theoretical interpretations include analytical \cite{Kazantsev1968}-type models, models with turbulent anisotropy induced by a mean velocity shear \citep[``shear-current'' effect,][]{Rogachevskii2003}, and other ideas \citep[for a review, see][]{Schekochihin2007}.

It is important to emphasize that the local dynamo mechanism in realistic solar conditions is different from both these cases. While the mean kinetic helicity on local dynamo scales is very small (zero in our model), strong stratification and highly anisotropic near-surface turbulence lead to strong local helical and shearing flows, particularly in downdrafts located in the intergranular lanes. Our simulations show that these flows are the primary source of the local dynamo in quiet-Sun regions.

Observational techniques are now able to resolve flows and magnetic fields associated with magnetic flux emergence events on very small subgranular scales \citep[e.g.][]{Centeno2007,OrozcoSuarez2008,MartinezGonzalez2009,Thornton2011,VargasDominguez2014}, which probably reflect the local dynamo at work in shallow subsurface layers. In addition, resolving magnetic features though observations on smaller and smaller scales has renewed interest in the small-scale magnetic field generation problem and raised a question about the existence of unresolved (or `hidden') magnetic flux in the quiet Sun \citep[e.g.,][]{Stenflo1982,Stenflo2012,TrujilloBueno2004,Shchukina2011}.

Theories of small-scale (or turbulent) dynamo processes have been intensively studied in the past. In particular, the similarity between the MHD induction and vorticity equations \citep{Batchelor1950} initiated discussions about magnetic field amplification due to turbulent flows. By analogy with vortices when stretching makes them stronger, twisting and stretching of magnetic field lines can amplify the field strength. Thus, repeating stretching, twisting, and folding of magnetic field lines can provide a local increase in the magnetic flux and its recirculation in a dynamo process \citep{Vainshtein1972b}.

Previous studies were mostly done with direct numerical simulations (DNS) performed for artificially forced turbulent flows \citep[e.g.,][]{Meneguzzi1989,Schekochihin2004,Schekochihin2005} and also for convectively driven flows for a wide range of parameters \citep[e.g.,][]{Nordlund1992,Cattaneo1999,Brandenburg2012}. In addition, recent `realistic'-type radiative MHD simulations have reproduced the solar surface conditions with a high degree of realism, and demonstrated that magnetic fields can be quickly amplified by local dynamos in the upper convection zone from a very weak ($\sim 10^{-2}$~G) seed field, leading to more than 1~kG magnetic elements \citep{Vogler2007,Kitiashvili2013b,Rempel2014}.

In this paper we present new results of 3D radiative MHD simulations of the local solar dynamo for various seed field strengths, from $10^{-6}$ to $10^{-2}$~G (in 5 different  simulation runs), and investigate the development and properties of the dynamo process in the realistic regime of solar magnetoconvection, when the turbulent magnetic Prandtl number is not prescribed a priori as a constant. Our goal is to investigate the primary physical mechanism of the magnetic field generation by local dynamo in the realistic solar conditions. A detailed statistical study and comparison are planned for a future paper.

Numerical simulations of local dynamos on the Sun have two specific features that must be taken into account: 1) the solar turbulence is driven by near-surface convective flows in a highly stratified medium; 2) the hydrodynamic and magnetic Reynolds numbers are so high that direct numerical simulations (DNS) with fully resolved turbulent scales are not possible.

The first feature requires accurate modeling of solar conditions including radiative transfer with realistic opacities and a realistic equation of state. This means that direct numerical simulations of flows with artificially forced turbulence have limited applicability to the solar dynamo problem. For our simulations, we use a radiative MHD code (``SolarBox") specifically developed for modeling solar conditions with high degree of realism. The basic formulation is presented in Section 2.1.

The second feature of solar simulations requires implementation of sub-grid scale (SGS) turbulence models. Previous solar magnetoconvection simulations \citep[e.g.][]{Stein2003b,Vogler2007,Rempel2014} used either artificial or numerical viscosity to model, explicitly or implicitly, dissipation due to turbulence at unresolved scales. In this work, we use a Large-Eddy Simulation (LES) approach implemented in the SolarBox code. This approach is widely used in engineering applications and tested with laboratory experiments and is still a topic of extensive study. The SGS models adopted in our solar simulations are developed and tested by using DNS MHD simulations including a fluctuating dynamo driven by forced turbulence. This approach provides a verification of the LES models. Some initial results of this project are described by \cite{Balarac2010}, and the LES models which currently provide the best performance (Balarac, private communication) are described in Section 2.1.

Perhaps the most significant implication of the LES approach for dynamo modeling is that in our simulations the characteristic turbulent dissipation parameters are not prescribed as constants, but modeled according to the local flow and magnetic field properties. This means that the critical non-dimensional parameters, such as the turbulent magnetic Prandtl number, are not prescribed but change with local conditions.

In Section 3, for the first time, we describe the physical mechanism of individual magnetic field generation events in terms of local plasma dynamics. We find a strong correlation between such events and local helical flows and vortex tubes. It is commonly believed that the flow helicity induced by rotation is responsible for the generation of the solar global (poloidal) magnetic field. No rotation is included in our simulations, but, because of the strong stratification of the turbulent surface convection, vortex tubes with helical motion are generated, mostly in high-speed (near sonic)  downdrafts in the intergranular lanes. The mechanism of vortex tube generation, dynamics, and associated atmospheric effects were described in our previous papers \citep{Kitiashvili2012b,Kitiashvili2012,Kitiashvili2013b,Kitiashvili2013a}. We show here that the local turbulent flow helicity in the solar magnetoconvection regime  plays a key role in the small-scale dynamo process.

In Section 4, we present our results on the statistical relationship between dynamo properties and the local turbulent magnetic Prandtl number. Such investigations have not been performed before. Also, previous investigations were limited to descriptions of stationary dynamo states.

In Section 5, we investigate the effects of dynamo-generated magnetic fields on the turbulent properties of solar magnetoconvection ("back reaction" of magnetic fields). A surprising result is that the resulting kinetic energy spectrum is closer to the Kolmogorov law than the initial spectrum without magnetic field. We discuss the evolution of the kinetic and magnetic energy spectra from seed-field initiation to the dynamo stationary state and also the variation of the spectra with depth. This allows us to investigate the process of energy transfer in the dynamo mechanism.

In Section 6, we describe the development of local bipolar magnetic structures that probably play an important role in the observed `salt-and-pepper' structure of the quiet-Sun magnetograms. Another significant new result is that the local dynamo can produce relatively long-lived compact magnetic structures. Their formation is related to the amplification of shearing and twisting flows caused by growing electric currents, and thus magnetic forces, around magnetic elements.

In Section 7, we discuss the topological properties of dynamo-generated magnetic fields in the solar atmosphere and show that they form a canopy-like structure consisting of small-scale magnetic loops. Such a topology results in a rapid change with height in the solar atmosphere of the ratio between the vertical and horizontal field components and can explain the discrepancies in the observed ratios \citep[e.g.][]{Danilovic2010}. The mean strength of the dynamo-generated magnetic fields corresponds to the observed values from Hinode \citep{Lites2008}. In Section 8, we summarize the principal results and discuss their implications for our understanding of solar magnetism, as well as future plans.

\section{Computational setup}
\subsection{3D radiative MHD `SolarBox' code}
To perform the local dynamo simulations we use the 3D radiative MHD code `SolarBox', developed at the NASA Ames Research Center by A. Wray and N. N. Mansour. The code is based on a LES (Large Eddy Simulation) formulation for compressible flow and includes a fully coupled radiation solver, in which local thermodynamic equilibrium is assumed. Radiative transfer between fluid elements is calculated using a 3D multi-spectral-bin method with long characteristics. For the initial conditions we use a standard solar model of the interior structure and lower atmosphere. The code has been carefully tested and was previously used for studying the excitation of solar acoustic oscillations by turbulent convection in the upper convection zone \citep{Jacoutot2008a,Jacoutot2008} and several other problems of solar dynamics and MHD \citep[e.g.][]{Kitiashvili2010,Kitiashvili2011,Kitiashvili2013a}.

We solve the grid-cell averaged equations of the conservation of mass (\ref{mass}), momentum (\ref{mom}), energy (\ref{energy}), and magnetic flux (\ref{eqB}):
\begin{equation} \label{mass}
\frac{\partial \rho}{ \partial t}+\left(\rho u_i\right)_{,i}=0,
\end{equation}
\begin{equation} \label{mom}
\frac{\partial \rho u_i}{ \partial t}+\left(\rho u_i u_j +
(P_{ij}+\rho\tau_{ij})\right)_{,j}=-\rho\phi_{,i},
\end{equation}
\begin{equation} \label{energy}
\frac{\partial E}{ \partial t}+
\left[E u_i + (P_{ij}+\rho\tau_{ij})u_j-(\kappa+\kappa^t) T_{,i}
+{\left(\frac{c}{4\pi}\right)}^2 \frac{1}{\sigma+\sigma^t}\left(B_{i,j}-B_{j,i} \right)B_j+
F_i^{rad}\right]_{,i}=0,
\end{equation}
\begin{equation} \label{eqB}
\frac{\partial B_i}{ \partial t}+\left[ u_j B_i - u_iB_j
-\frac{c^2}{4\pi(\sigma+\sigma^t)} \left(B_{i,j}-B_{j,i} \right)\right]_{,j}=0,
\end{equation}
where $\rho$ is the averaged mass density, $u_i$ is the Favre-averaged (density-weighted) velocity, $B_i$ is the magnetic field,
and $E$ is the averaged total energy density $E=\frac{1}{2}\rho u_i u_i + \rho e + \rho\phi +\frac{1}{8\pi}B_iB_i$, where $\phi$ is the gravitational potential and $e$ is the Favre-averaged internal energy density per unit mass. $F_i^{rad}$ is the radiative flux, which is calculated by solving the radiative transfer equation, and $P_{ij}$ is the averaged stress tensor $P_{ij}=\left(p+\frac{2}{3}\mu
u_{k,k}+\frac{1}{8\pi}B_kB_k\right)\delta_{ij}-\mu\left(u_{i,j}+u_{j,i}\right)-\frac{1}{4\pi}B_iB_j$, where $\mu$ is the dynamic viscosity. Gas pressure $p$ is a function of $e$ and $\rho$ calculated through a tabulated equation of state \citep{Rogers1996};  $\tau_{ij}$ is the Reynolds stress, $\kappa$ is the molecular thermal conductivity; $\kappa^t$ is the turbulent thermal conductivity,  $\sigma$ is the molecular electrical conductivity, and $\sigma^t=c^2/4\pi \eta_m^t$ is the turbulent electrical conductivity in Gaussian units. The magnetic diffusivity, $\eta_m^t$, is calculated using the subgrid-scale (sgs) model  \citep{Theobald1994,Balarac2010}: $\eta_m^t=a\triangle^2|\nabla\times{\rm\bar B}|/\sqrt{\rho}$, where $a$ is a constant ($a=2$), $\triangle$ is the grid spacing, and overbars denote the resolved field. The turbulent heat conductivity $\kappa^t$ is given in terms of the turbulent thermal Prandtl number, $Pr^t$,  as $\kappa^t=\rho c_p \nu^t/Pr^t$, where $\nu^t=\mu^t/\rho$. We assume that $Pr^t$  is unity.

Currently it is impossible to achieve a realistic Reynolds number in direct numerical simulations of solar MHD phenomena; therefore modeling of the dynamical properties of solar convection is performed through implementation of subgrid-scale  LES (Large Eddy Simulation) turbulence models. These models create a realistic turbulent Reynolds number at which the simulations can be accurate and so provide a better representation of small-scale motions \citep[e.g.][]{Zhou2014}. Here we used a Smagorinsky eddy-viscosity model \citep{Smagorinsky1963}, in which the compressible Reynolds stresses were calculated in the form \citep{Moin1991}: $\tau_{ij}=-2C_S\triangle^2|S|(S_{i,j}-u_{k,k}\delta_{ij}/3)+2C_C\triangle^2|S|^2\delta_{ij}/3$, where the Smagorinsky coefficients are $C_S=C_C=0.001$; $S_{ij}$ is the large-scale stress tensor.

The simulation results presented here were obtained for a computational domain of $6.4\times6.4\times6.2$~Mm, which includes 1-Mm of the lower atmosphere. The grid-size is 12.5~km in the horizontal direction. Above the solar surface the vertical grid size is constant (12~km) but stretched out in the deeper layers. The lateral boundary conditions are periodic. The top boundary is open for mass, momentum, and energy fluxes, and also for the radiation flux.
The top boundary is handled through characteristic boundary conditions; hence fluxes of all quantities can occur through this boundary in either direction. The bottom boundary is handled as follows: 1) the vertical velocity component is set to zero;
2) horizontally uniform numerical fluxes are created at the bottom boundary to compensate for any net inward or outward flux through the top boundary, for mass, momentum, energy, and B-field, except that the radiative energy flux through the top is not counted in this process; 3) an energy flux equal to the stellar luminosity is introduced at the bottom boundary, also uniformly in the horizontal dimensions.
As a result, all conserved quantities retain constant integrals over the computational volume, except that the total
energy is allowed to vary due to radiative loss and the steady bottom-boundary flux that mimics energy flow from the core.  This seems to us to be a physically natural treatment of the energy, and it also serves as a check on the radiative transfer calculation.  We have verified that, over a sufficiently long time interval, the constant power input at the bottom boundary comes to be compensated by the radiative flux through the top boundary, resulting in a total energy that oscillates randomly and with very small amplitude around a constant value.
In summary: 1) mass, momentum, energy, and B-field can advect and diffuse through the top boundary; 2) these losses or gains are compensated at the bottom boundary; 3) power to satisfy the stellar luminosity is introduced at the bottom, and radiation through the top quickly comes to balance this input within a few percent and eventually balances it exactly in time average.

Compared to early magnetoconvection simulations by \cite{Stein2003b} and later work by \cite{Vogler2007} and \cite{Rempel2014}, our simulation domain is significantly deeper: 5.2 Mm vs 2.5 Mm of \cite{Stein2003b}, 0.9 Mm of \cite{Vogler2007}, and 2.3 Mm of \cite{Rempel2014} in his high-resolution runs. The deeper domain reduces the effects of the bottom boundary conditions. The simulation results presented in Section 3 show that the primary region of magnetic field generation and energy exchange is located in the upper 1 Mm layer, but slowly extends into deeper layers. Therefore, it is important that the computational domain has a significant depth.

\subsection{Initial conditions}
Local dynamo action is a complicated interaction of magnetic fields and highly turbulent flows on small scales. In this paper, the dynamo simulation is started by adding a very weak seed field to a hydrodynamic simulation of fully developed solar convection. After introducing the `seed' field, \underline{no additional magnetic flux is introduced into the domain}.  To investigate the effects of the initial seed-field properties, we consider 5 cases of magnetic field initialization (Table~\ref{tab:cases}). In these cases the initial seed magnetic field has various distributions: ($A$) uniform vertical field, ($B$) checkerboard-like, alternating polarity vertical field patterns, and ($C - E$) a randomly distributed net of magnetic field with amplitude $\pm 10^{-2} - 10^{-6}$~G (white noise for all field components). In case $B$ the checkerboard structure has a period of magnetic field variations of 100~km, in order to mix opposite-polarity patches in the intergranular lanes. The hydrodynamic conditions at the time of magnetic field initialization are exactly the same for cases $B - E$.

\begin{table}[h]
 \begin{center}
 \caption{Properties of the seed magnetic field. \label{tab:cases}}
    \begin{tabular}{|c|c|c|}
        \hline
         Cases & Magnetic field & Initial field \\
          & strength, G & configuration \\
        \hline
            $A$ & $10^{-2}$ & vertical \\
            $B$ & $10^{-2}$ & checkerboard \\
            $C$ & $10^{-2}$ & white noise \\
            $D$ & $10^{-4}$ & white noise \\
            $E$ & $10^{-6}$ & white noise \\
        \hline
    \end{tabular}
 \end{center}
\end{table}

\section{Generation of small-scale fields by turbulent plasma}
Magnetic field amplification in turbulent solar convection can roughly be divided into three basic mechanisms: 1) magnetic field concentration due to converging flows \citep[e.g.][]{Nordlund1983}, 2) convective collapse \citep{Parker1978,Spruit1979}, and 3) dynamo processes driven by helical or shearing motions \citep[e.g.][]{Batchelor1950,Vainshtein1972b,Brandenburg1995a}. The turbulent nature of the photospheric layers, where the intense radiative cooling drives downward convective motions, makes it impossible to separate different sources of the locally growing magnetic energy \citep{Spruit1984a}, and therefore the magnetic field amplification can be discussed only in terms of a dominant mechanism.

To avoid confusion in separating dynamo and non-dynamo field amplification, no additional magnetic flux was introduced into the computational domain after the `seed'-field initialization. This provides an important test for the dynamo action: if there were no dynamo process then local magnetic patches would only be formed due to converging flows in the intergranular lanes, and then convective downdrafts would transport these patches into deeper layers, from which a part of the flux would be recycled by turbulent motions. In contrast, with small-scale dynamo action magnetic field patches are continuously formed, resulting in a rapid growth of magnetic energy.

\begin{figure}
\begin{center}
\includegraphics[scale=0.8]{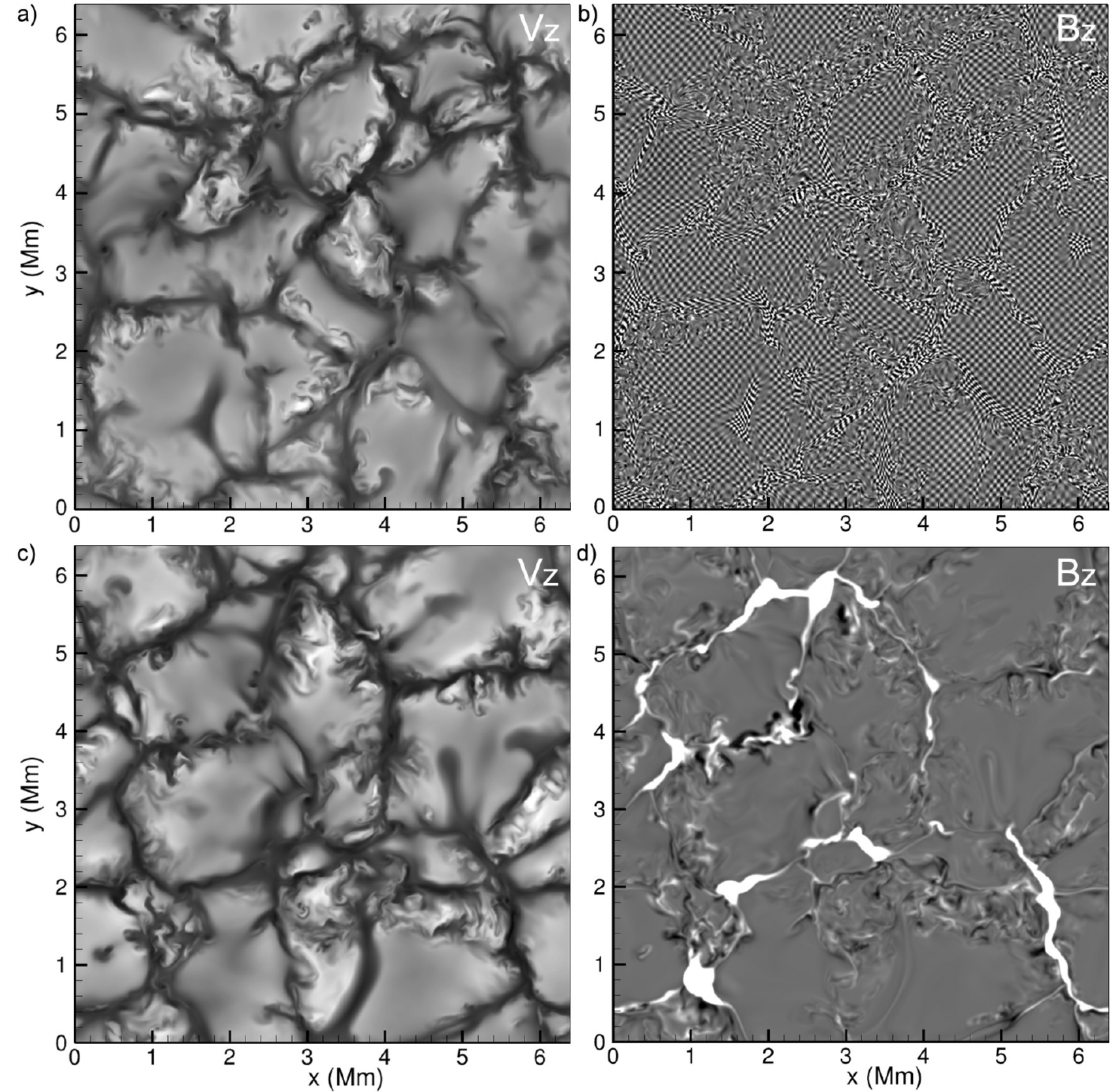}
\end{center}
\caption{Deformation of the initial checkerboard-distributed magnetic field (panel $b$) for the case $B$ (${\rm B}_{0z}=\pm 10^{-2}$~G) due to the surrounding turbulent convection (panel $a$) at the photosphere for $t=30$~sec after field initialization. Black-white patterns correspond to opposite-polarity magnetic fields, saturated in this image at $\pm 10^{-2}$~G. Panels $c$) and $d$) show distributions of the vertical velocity and magnetic field respectively in the developed state, $\sim 4$~hours after the field initialization, at $z = 0$ and correspond to the last frame of the 3D movie showing volume rendering of the magnetic field strength (see supplementary materials). In panel $d$) the magnetic field image is saturated at $\pm 100$~G.  \label{fig:checker}}
\end{figure}

We note that during very early times of the `dynamo' runs (shorter than one overturning time) the evolution of the `seed'-field magnetic elements behaves similarly to corks in distributed turbulent flows: they tend to collect in the intergranular lanes where the convective flows converge. Figure~\ref{fig:checker} shows the distribution of the vertical velocity (panel $a$) and magnetic field (panel $b$) in the photosphere shortly after introducing the `seed' magnetic field as a checkerboard structure (case $B$, see Table~\ref{tab:cases}). Because the weak field mostly follows the turbulent flow, the regular field distribution becomes deformed. During this stage local magnetic field amplification is primarily caused through compression by converging flows in the intergranular lanes. Shortly after this initial phase we see the start of the dynamo process: small-scale helical motions in the intergranular lanes drag magnetic field lines and stretch them, and can even reverse the initial local polarity of the field.  The dynamo process becomes stronger with time as turbulent flows on larger scales become involved. The mean magnetic energy density becomes saturated after 7 -- 8 solar hours. Snapshots of the vertical velocity and magnetic field in the developed state at $z = 0$ are shown in Fig.~\ref{fig:checker}$c$, $d$. The maximum magnetic field magnitude reaches 2~kG, and the photospheric mean magnitude reaches $\sim 20$~G (Fig.~\ref{fig:hist1}$c$).

The dynamics of the small-scale dynamo is illustrated in the movie included in the supplementary materials, in which the semi-transparent color scale corresponds to the magnetic field strength. The color scale is saturated at 500~G, but the field strength can reach more than 2000~G at the photosphere. After the seed-field initialization, the first signatures of the amplified field (light green-blue diffuse structures) appear above the photosphere due to the short dynamical time-scales there.  Shortly thereafter the first magnetic patches appear at the photosphere. The appearance of opposite-polarity patches reveals a `salt-and-pepper' distribution of magnetic fields (Fig.~\ref{fig:checker}$d$). The patches evolve rather chaotically, but nevertheless the maximum magnetic field strength grows continuously until it finally saturates. These magnetic elements are highly twisted and show a very dynamic and non-stationary behavior, and none of them survive for long.

Because the local dynamo is driven by turbulent motions that cause magnetic field amplification due to the stretching and twisting of field lines, it is natural to consider the relationship between turbulent plasma motions and the local evolution of the magnetic field. For instance, a relationship between the dynamo process and vortex motions produced by the Coriolis force in a rotating stratified medium was demonstrated numerically by \cite{Nordlund1992}. Our simulations  reveal small-scale vortex tubes in the intergranular lanes; however, no external rotation has been introduced. These vortex tubes are a result of overturning convection and Kelvin-Helmholtz type instabilities \citep{Kitiashvili2012}. They play a significant role in the small-scale dynamo mechanism.

To quantify the apparent close relation between the appearance of magnetic patches and local twisting motions observed in our simulations, we calculate the kinetic helicity density, $H=\mathbf{v}\cdot(\nabla\times\mathbf{v})$, and investigate its correlation with  the time derivative of the magnetic energy density. A similar relationship between the kinetic helicity density and local amplification of magnetic flux \citep[analogous to the shear-twist-fold dynamo of][]{Vainshtein1972}  was previously noticed by \cite{Brandenburg1996}. In their simulations the vorticity was induced by rotating the simulation box. In our case without global rotation, the mechanism of vorticity generation due to shearing instabilities and compression has been discussed in our previous paper \citep{Kitiashvili2012b}.

\begin{figure}
\begin{center}
\includegraphics[scale=1.2]{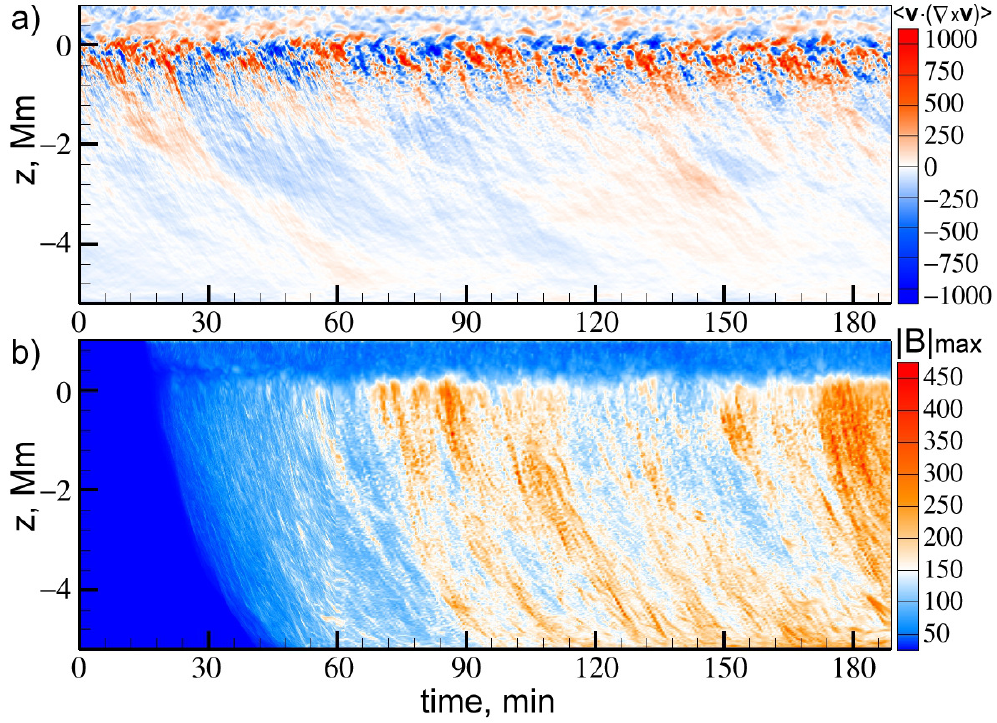}
\end{center}
\caption{Time-depth evolution of the mean helicity density, $<H>=<\mathbf{v}\cdot(\nabla\times\mathbf{v})>$ (panel $a$), and strongest magnetic field strength in a  horizontal plane (panel $b$) for case $E$ (Table 1) with an initial seed field of $10^{-6}$~G. \label{fig:time-depth}}
\end{figure}

Vertical profiles of the {\it rms} velocity and the kinetic helicity density, $H=\mathbf{v}\cdot(\nabla\times\mathbf{v})$, displayed in Figure~\ref{fig:time}, show that the strongest turbulent flows occupy a relatively thin 1-Mm deep layer just below the solar surface. Despite the similarity of the vertical profiles for velocity and helicity, the kinetic helicity density is more concentrated in the upper 1-Mm subsurface layer. Our previous studies of helical motions (vortex tubes) in the near-surface layers \citep{Kitiashvili2012} showed that the lifetime of the vertically oriented vortex tubes can be much longer (up to 1~hour) than the granulation lifetime ($\sim 10$~min). The vortex tubes are characterized by strong flows reaching sonic speeds, by a sharp decrease of the gas pressure, and by intense radiative cooling. The strong horizontal and vertical vortex motions are able to attract small magnetic patches, amplify magnetic field by stretching the field lines, and, via downdrafts, transport the field into deeper layers, where the field can be further amplified by compression (Fig.~\ref{fig:time-depth}).

To investigate the statistical relationship between twisting motions and magnetic field generation, we calculate cross-correlations of the squared magnetic field strength, $B^2$, with the kinetic helicity density and with the magnitude of the shear stress, $|S|=(2S_{ij}S_{ji})^{1/2}$ \citep{Moin1991} (Fig.~\ref{fig:cross-corrHelU-B2}). The result is that the distributed magnetic field patches have better correlation with the kinetic helicity density (Fig.~\ref{fig:cross-corrHelU-B2}$a$) than with the magnitude of shear stress (Fig.~\ref{fig:cross-corrHelU-B2}$b$). These correlation functions decrease with depth more slowly for the kinetic helicity density due to transport of the magnetic field by helical downflows (Fig.~\ref{fig:time-depth}).
Thus, we conclude that the small-scale dynamo under solar conditions primary develops due to local fluctuations of the kinetic helicity density, and that shearing fluctuations might also contribute to the small-scale dynamo.

\begin{figure}
\begin{center}
\includegraphics[scale=0.85]{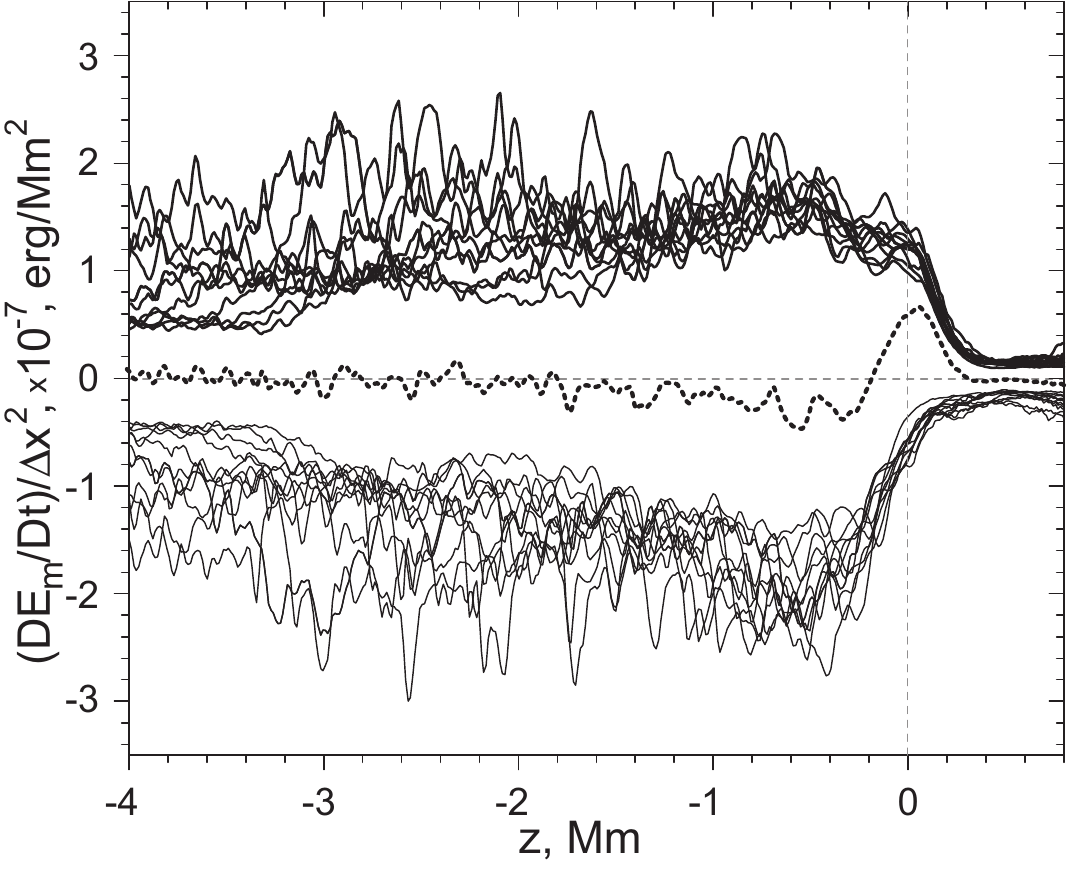}
\end{center}
\caption{Profiles of the time-derivative of the mean magnetic energy density divided by the total area occupied
by positive and negative values, respectively, ${\rm A=\Delta x\Delta y}$, as a function of depth, plotted separately for the growth (thick  curves) and decay (thin curves). Each curve is plotted for data taken every 10 minutes, (case $E$).  The dotted curve shows the derivative of  the total mean magnetic energy density averaged over time. \label{fig:DEm-depth}}
\end{figure}

Because magnetic field generation and dissipation are highly inhomogeneous and depth-dependent, we analyze areas where the magnetic energy grows and dissipates (Fig.~\ref{fig:DEm-depth}). We find that the time derivative of the magnetic energy density is small above the photosphere, while at the photosphere the growth rate of the magnetic energy is faster than dissipation. In the subsurface layers the rates of growth and dissipation of magnetic energy are very similar. In layers deeper than 1~Mm, the rates of energy variation slowly decrease and are characterized by increasing amplitude of the local energy fluctuations. We plan to investigate the atmospheric effects in a separate paper.

\section{Turbulent magnetic Prandtl number and generation of small-scale fields}
Magnetic field amplification due to swirling and shearing flows has been investigated by many authors \citep[see e.g.][]{Batchelor1950,Vainshtein1972b,Nordlund1992,Petrovay1993a,Childress1995,Brandenburg1996,Brandenburg2012}.
Stretching, twisting, and folding of magnetic field lines induce growth of the magnetic flux under certain plasma conditions, such as a sufficiently high magnetic Reynolds number. The magnetic Prandtl number, which is the ratio of the diffusion coefficients of momentum and magnetic field, also plays a critical role in dynamo conditions \citep{Nordlund1992,Vogler2007}. Previously it was shown that magnetic field can be easily amplified for large magnetic Prandtl numbers ($\geq 1$),  \citep[e.g. see][]{Pao1963,Meneguzzi1989,Brandenburg1996,Schekochihin2004}. However, estimates of the magnetic Prandtl number for the Sun, based on molecular diffusivities, vary with depth from $\sim 10^{-1} - 10^{-2}$ near the base of the convective zone to $\sim 10^{-5}$ at the surface \citep{Rieutord2010}. This suggests that the dynamo cannot operate efficiently since, under these conditions, magnetic diffusion is much higher than momentum diffusion. Recently, however, it was found that dynamos can work for small magnetic Prandtl numbers \citep[e.g.][]{Ponty2004,Schekochihin2005,Iskakov2007} and that the transition between the regimes where magnetic field is either amplified or diffused by turbulent plasma motions is described by a critical magnetic Prandtl number, $Pr^c_m$, below which the dynamo processes do not occur. For theoretical discussions of this problem in the case of incompressible artificially forced turbulence, see papers by \cite{Rogachevskii1997,Schekochihin2004a,Schekochihin2007,Tobias2011a}.

In our numerical simulations, the molecular scales are unresolved, and LES turbulence models are used to estimate the plasma dynamics and diffusivities due to motion on sub-grid scales. Therefore, in our simulations the magnetic Prandtl number reflects the properties of plasma and magnetic fields only on turbulent scales, and is defined as the ratio of the local turbulent plasma and magnetic diffusion coefficients ($Pr^t_m=\nu^t/\eta_m^t$). Figure~\ref{fig:prm} illustrates typical snapshots of the vertical velocity, magnetic field strength, electric current density, the time-derivative of magnetic energy density, the turbulent magnetic Prandtl number, and the turbulent magnetic diffusivity in a fraction of the simulation domain at the photosphere ($z = 0$). The simulations reveal concentration of the magnetic field in the intergranular lanes and highly inhomogeneous time-evolution of the magnetic energy on scales smaller than the width of the lanes. In particular, we find that the magnetic energy grows mostly at the edge of granules and at the periphery of turbulent vortex tubes, where shear motions are strongest, whereas the decrease of magnetic energy mostly occurs in areas where strong downflows transport magnetic field into the deeper layers. This inhomogeneity is also reflected in the distributions of turbulent magnetic Prandtl number $Pr_m^t$ and magnetic diffusivity $\eta_m^t$. Small values of $Pr_m^t$ are mostly associated with areas where shearing flows are present and the magnetic diffusivity has its largest values.

\begin{figure}
\begin{center}
\includegraphics[scale=1.2]{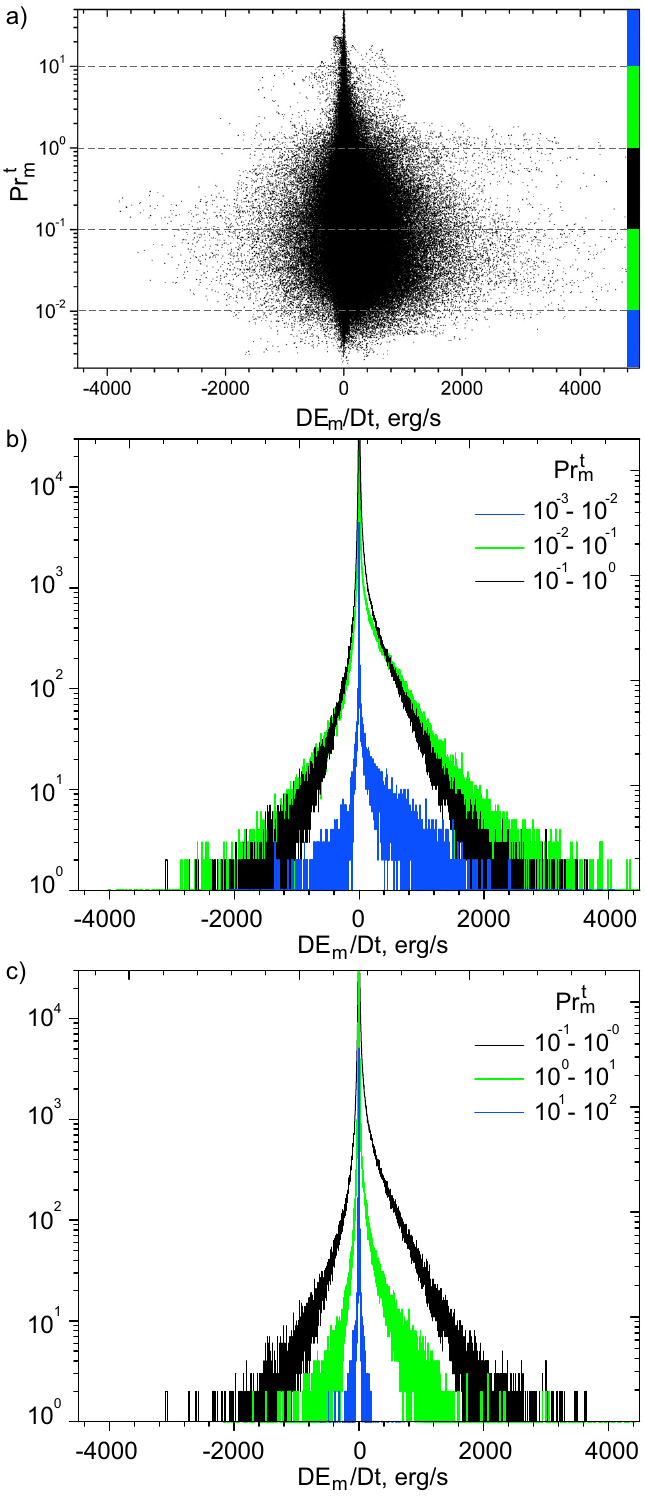}
\end{center}
\caption{$a$) Statistical distribution of the $Pr_m^t$ vs. Lagrangian derivative ${\rm DE_m/Dt}$. $b$) and $c$) histograms of ${\rm DE_m/Dt}$ for the various ranges of the magnetic Prandtl number indicated in the panels.  \label{fig:hist2}}
\end{figure}

A statistical distribution of the turbulent magnetic Prandtl number ($Pr^t_m$) for a simulation of 1-hour of solar time is shown in Figures~\ref{fig:hist1} and~\ref{fig:hist2}. According to our simulation results, $Pr^t_m$ can vary by several orders of magnitude, and most values are in the range of  $0.1 -  1$ (Fig.~\ref{fig:hist1}$b$). The local growth and decay of the magnetic energy density correlate with turbulent Prandtl values $Pr^t_m \sim 10^{-2} -  1$ (Fig.~\ref{fig:hist2}$a$). The distribution of the magnetic energy density rate of change for different ranges of the magnetic Prandtl number (Fig.~\ref{fig:hist2}) shows a distinct asymmetry between the decay and amplification of magnetic field, in particular, for smaller $Pr_m^t$. This is clear evidence that the local dynamo processes are particularly efficient in areas characterized by low magnetic Prandtl number.

\section{Effect of small-scale dynamo action on the turbulent properties of solar magnetoconvection}
The quiet Sun regions are characterized by relatively weak mean magnetic flux on the solar surface, but this flux does affect the turbulent properties of convection on small scales, where magnetic field patches can partially suppress plasma motions.
For instance, comparison of the turbulent spectra of the photosphere, calculated from simulations with and without magnetic field, shows that the slope of the inertial range changes from $-11/5$ in the purely hydrodynamic case to $-5/3$ in the presence of magnetic fields \citep{Kitiashvili2013}. Thus, the dynamo-generated magnetic fields locally act on surrounding turbulent flows (the so-called `back reaction'). In the current simulations, which account for turbulent dynamics on sub-grid scales, we are able to capture the complicated interaction and energy exchange between the small-scale fields and flows.

The dynamo process produces magnetic patches with a wide range of scales of magnetic field distribution (Fig.~\ref{fig:hist1}$c$) and involves the redistribution of energy among different scales. Therefore, we consider the development of dynamo  as an evolution of the turbulent kinetic energy density spectra, which show a redistribution of energy through different scales during the transition from the `seed' field to a saturated magnetoconvective regime (Fig.~\ref{fig:power1}$a$). In particular, we find that the initial development of magnetic patches on the smallest scales increases small-scale electric currents and kinetic energy (red curve on panel $a$). Then the energy shift from small to larger scales changes the slope of the spectra so that they approach $k^{-5/3}$ (green curve). At the same time, the continuing amplification of magnetic fields at large wave numbers (small scales) causes suppression of the kinetic energy on these scales. Thus, the evolution of the kinetic energy-spectra during the development of the dynamo process can be interpreted as partly energy transport from smaller to larger scales and partly as a transformation of turbulent kinetic energy into magnetic energy. Note that, because of the relatively small size of the computational domain and occasional appearance of intermittent `bursts' of the local magnetic field generation \citep[similar to previously observed the bursts in DNS simulations by][]{Pratt2013}, variations of the energy spectra can be due to transient energy redistributions across turbulent scales (e.g., see the spectrum shown by blue curve in Fig.~\ref{fig:power1}$a$). Curiously, in the  saturation phase (Fig.~\ref{fig:power1}$b$) the kinetic energy density spectrum at the photosphere is of the Kolmogorov type ($k^{-5/3}$), whereas the results by \cite{Rempel2014} do not reproduce this slope. Also, the kinetic energy density in Rempel's simulations at k=10 Mm$^{-1}$ is an order of magnitude higher, and at k=100 Mm$^{-1}$ is almost two orders of magnitude higher compared to our simulations. We do not know the reason for this difference, but perhaps the higher kinetic energy density is a result of the shallower simulation domain (2.4~Mm) in Rempel's simulations. The characteristic scale of the solar granulation at the photosphere is in agreement with both observations and Rempel's simulations (note, the definition of the horizontal scale $f=k_h/2\pi$ in his paper).
In deeper layers the slope of the spectrum in the inertial range decreases in magnitude: the power law is closer to $k^{-6/5}$ at a depth of 1 Mm, and the slope is even smaller in deeper layers.

\begin{figure}
\begin{center}
\includegraphics[scale=0.8]{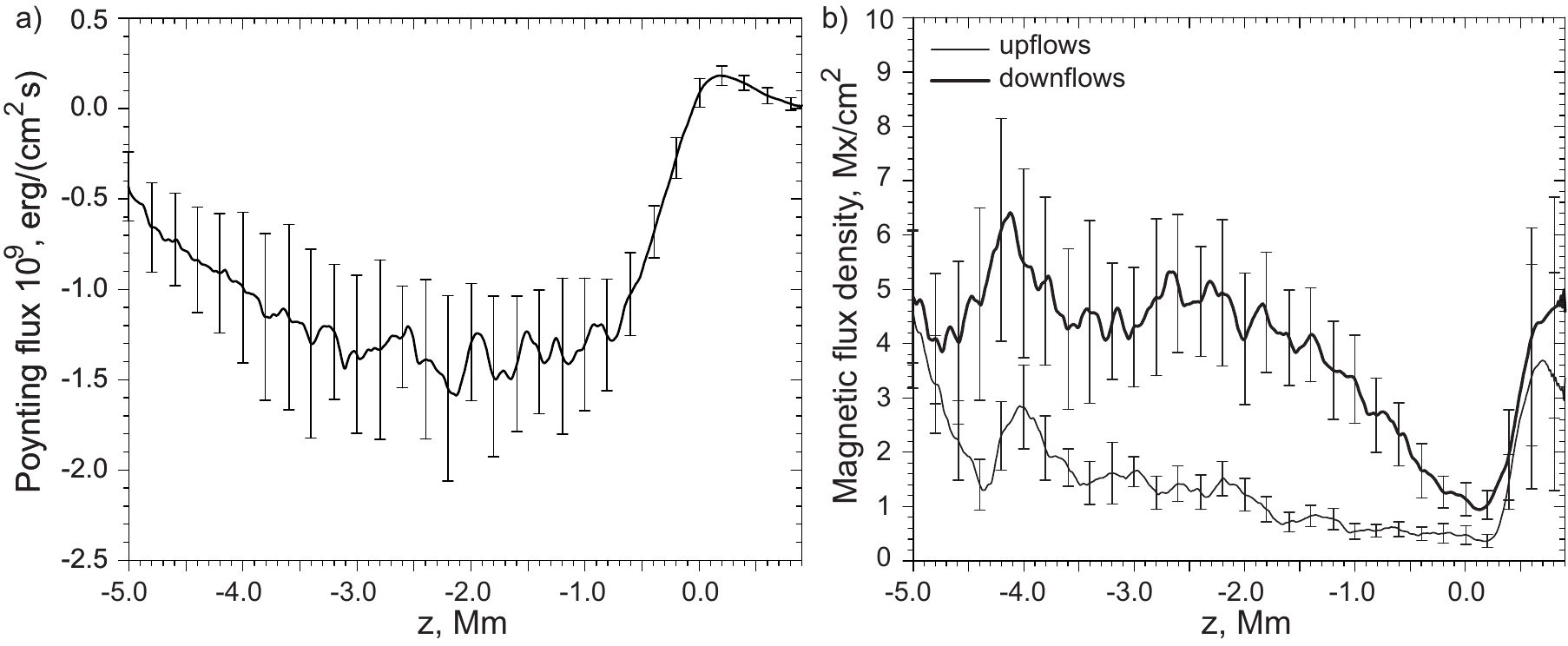}
\end{center}
\caption{ $a$) Mean vertical Poynting flux as a function of depth; $b$) mean horizontal magnetic fluxes in upflows (think curve) and in downflows (thick curve).  All properties are averaged over 1 hour. Error bars shows the standard deviation. \label{fig:flux}}
\end{figure}

Figure~\ref{fig:power} shows time and depth variations of the power spectra of the magnetic energy density from 2 Mm below the solar surface up to the photosphere. We found significant variations of the magnetic energy spectrum during the first half-hour after the  `seed' field initialization, i.e., during the exponential growth phase.  Later in time, the shape of the spectrum does not change, and the spectral power shows a steady energy increase at all scales.
The slope of the magnetic energy spectra for small wavenumbers changes after field initialization. It reaches a value of 1/3 after $\sim 3 - 4$~hours in the subsurface layers and a bit earlier in the photosphere. Our simulation results show the existence of a $k^{-1}$ power law in the saturation phase only for depths of about of 2~Mm and below, where the distribution of magnetic field is more homogeneous. We do not reproduce the theoretical $k^{-11/3}$ law \citep{Golitsyn1960,Moffatt1961}. This possibly reflects deviations from the theoretical assumptions of magnetic field weakness, and thus the absence of magnetic field back-reaction, in previous models. Note, that \cite{Rempel2014} didn't see a clear indication of a power law for the magnetic energy in his simulations. Figure~\ref{fig:flux} shows the averaged over one hour vertical distribution of the Poynting flux and the horizontal magnetic flux density transported by upflows and downflows. The magnitude of the Poynting flux is similar to the values found in Rempel's simulations for the open boundary/zero field boundary conditions. However, in our simulations the Poynting flux is decreasing at depths below 3 Mm (Fig.~\ref{fig:flux}$a$) while in the Rempel's simulations it continues to increase. The reason for this difference is unclear, and deserves further detailed investigation. Most of the horizontal flux is transported downflows into the deeper convection zone (Fig.~\ref{fig:flux}$b$). The mean flux recirculating by upflows is several times smaller than the flux in downflows.

\section{Formation of small-scale bipolar structures}
Understanding the local small-scale magnetic field generation process is critical for studying complex solar MHD problems, such as the interaction of magnetic fields and flows through different scales. In addition to the statistical properties described in previous sections, it is important to consider this process in terms of local physical properties. Here we present two characteristic cases of local magnetic field amplification.

\begin{figure}
\begin{center}
\includegraphics[scale=1]{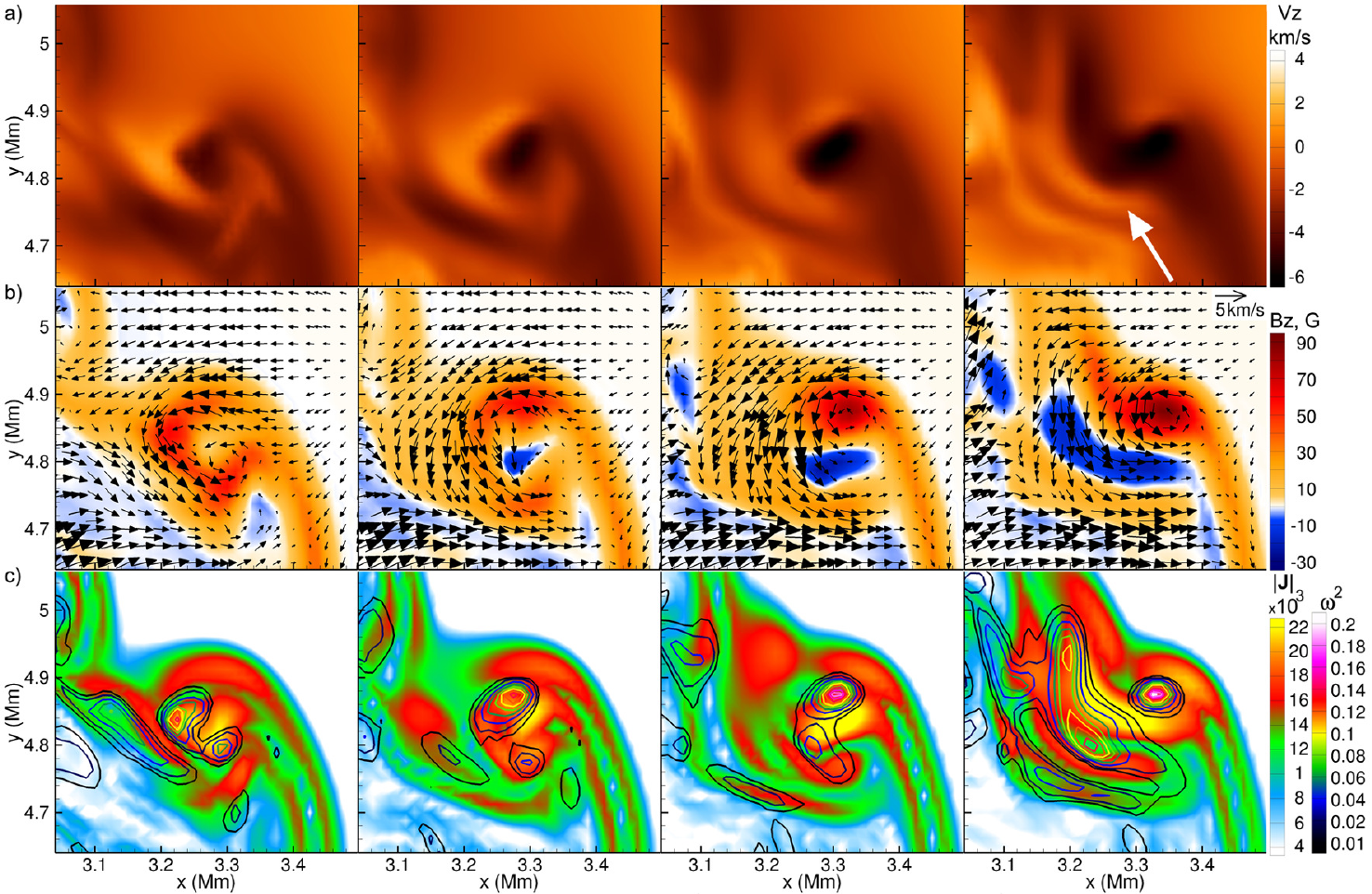}
\end{center}
\caption{Time-sequence with 15~sec cadence in a zoomed $400$~km $\times 400$~km region, where the magnetic field is generated by swirling turbulent flows: $a$) the vertical velocity in the photospheric layer ($z=0$); $b$) the vertical magnetic field evolution showing the development of small-scale magnetic elements with opposite polarity (bipolar magnetic structure); black arrows represent the horizontal velocity field; and c) the electric current density (background image) and the squared magnitude of vorticity ($\omega^2$, contour lines). This example corresponds to case~$A$, with the initial $10^{-2}$~G seed field. The white arrow points to a horizontal vortex tube discussed in the text. \label{fig:case1}}
\end{figure}

The first case, demonstrated in Fig.~\ref{fig:case1}, shows the  evolution of the vertical velocity, the vertical magnetic field (color background), the horizontal velocity field (arrows), the electric current density (color map), and enstrophy (contour lines) in a selected $400$~km $\times$ 400~km region of the photosphere with 15~sec cadence.

The development of a local bipolar magnetic structure is associated with strong ($\sim$5~km/s) swirling motions, on scales from $\sim 12-25$~km to $\sim 300-400$~km, stretching and twisting the magnetic field, which is also compressed by converging flows in the downdrafts. Such small-scale swirling motions in the intergranular lanes are typically associated with strong downflows, $\sim 6-8$~km/s  \citep{Kitiashvili2010,Kitiashvili2011}. The appearance of a bipolar magnetic structure (with a prominent negative (blue) polarity in Fig.~\ref{fig:case1}b) is also a result of swirling flows driven by a vortex tube oriented along the solar surface (indicated by the white arrow in Fig.~\ref{fig:case1}~$a$), similar to one previously described by \cite{steiner2010}. The horizontally oriented vortex tube captures magnetic field lines and drags them into the subsurface layers. The dynamics of the positive polarity patch is mostly related to the vertically oriented helical motions similar to the process described by \cite{Kitiashvili2010,Kitiashvili2011}  who showed that the process of spontaneous formation of stable magnetic structures consists of two basic steps: (1) formation of small-scale filamentary magnetic structures associated with concentrations of vorticity and whirlpool-type motions, and (2) merging of these structures due to vortex attraction, caused by converging downdrafts around magnetic concentration below the surface. Such complicated flow dynamics is caused by interaction of differently oriented vortex tubes, which, in fact, have orientations varying in space and only locally can be regarded as having `vertical' or 'horizontal' orientation \citep{Kitiashvili2012}. The generated electric current (Fig.~\ref{fig:case1}~$c$) shows a clear correlation with the swirling dynamics of convective flows. The strongest electric current density corresponds to areas on the periphery of helical motions or between the vortices, where stretching of the magnetic field lines is the strongest.

\begin{figure}
\begin{center}
\includegraphics[scale=0.8]{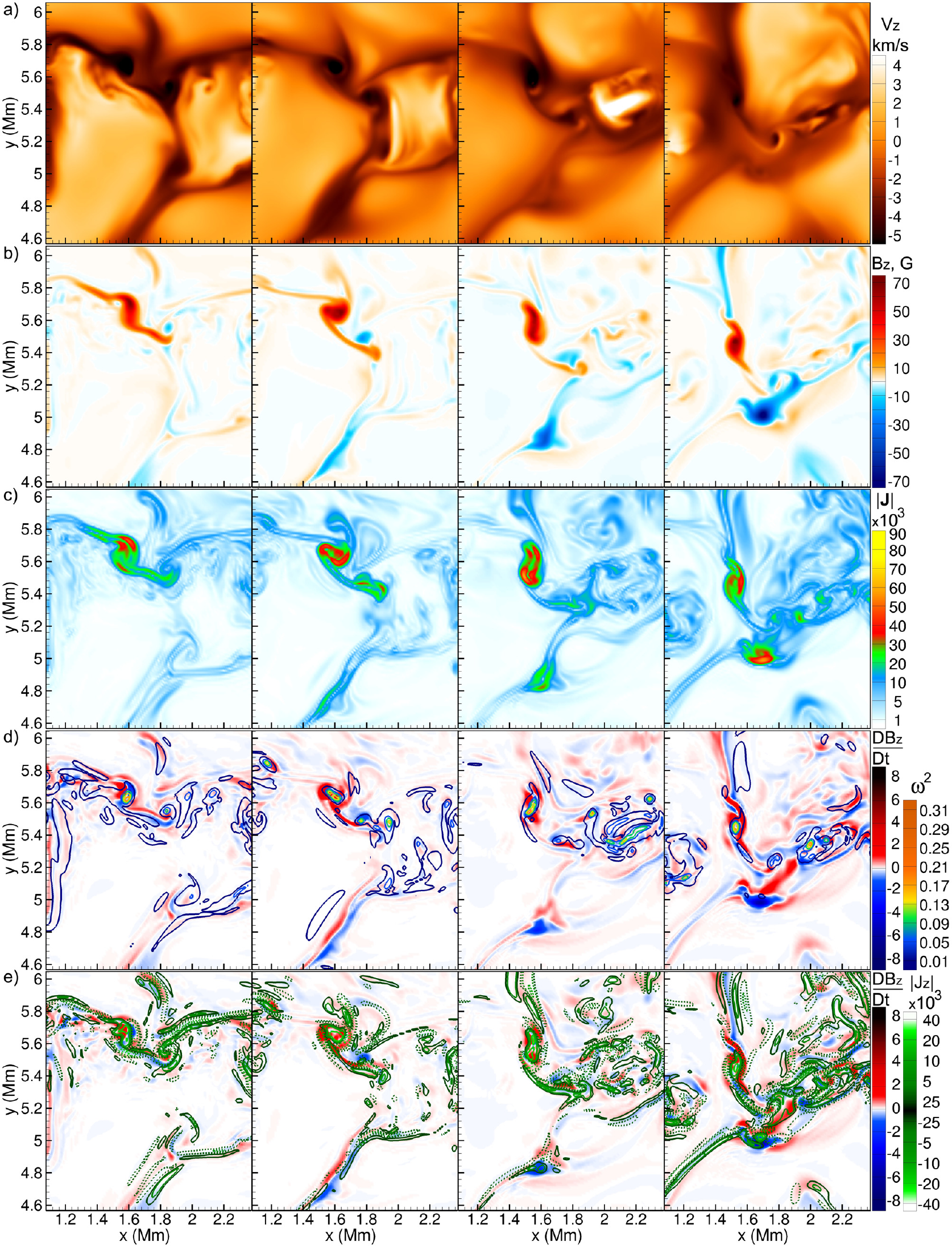}
\end{center}
\caption{Evolution of a bipolar magnetic structure in the photospheric layer ($z=0$) illustrated in a sequence of four images with cadence 90~sec for different parameters: $a$) vertical velocity; $b$) vertical magnetic field; $c$) magnitude of electric current density; $d$) and $e$) show the vertical magnetic field growth rate as a background red-blue image; in panel $d$) contours correspond to enstrophy and in panel $e$) contour lines show the vertical component of electric current (dashed curves for negative values). \label{fig:case3_2D}}
\end{figure}

Another example illustrates the development of a bipolar magnetic structure due to helical flows and shearing flows along an intergranular lane. Figure~\ref{fig:case3_2D} shows a time sequence of the vertical velocity, the vertical magnetic field, the magnitude of the electric current density, the time-derivative of the vertical magnetic field with overlaid contour lines of enstrophy, and the vertical component of the electric current density (dashed curves correspond to negative values). An interesting feature of this example is a different scenario for the development of magnetic elements of opposite polarity (determined by the sign of the vertical field component): the positive-polarity magnetic patch evolves following the `classical' scenario of field amplification due to helical flows described in the first example, whereas the negative-polarity patch starts forming in the intergranular lane mostly due to converging flows (magnetic collapse). The evolution of the magnetic elements of this bipolar structure is accompanied by locally growing electric current (Fig.~\ref{fig:case3_2D}~$c, d$). Comparison of the velocity, pattern, and time derivative of the vertical magnetic field shows that some swirling motions  induce magnetic field dissipation, probably due to scattering of the field lines, and often the field amplification takes place on the periphery of swirls or regions with shearing flows.

The distribution of vertical electric current density (Fig.~\ref{fig:case3_2D_E-current}) shows a strong correlation with the horizontal velocity of swirling flows in the subsurface layers. In deeper layers, the magnitude of the electric current increases, but its distribution becomes diffuse and does not show a clear association with the near-surface dynamics. The kinetic helicity density patterns show that the scale of the swirling motions increases with depth, from $\sim 50-60$~km at the photospheric layer (Fig.~\ref{fig:case3_2D_KinHel}) up to $~120$~km at a depth of 300~km below the photosphere (panel $d$). In the deeper layers, the scale of helical motions continues to increase (to larger than 150~km), but the distribution of kinetic helicity density becomes complicated and consists of opposite-sign helical flows, which however continue swirling together (Fig.~\ref{fig:case3_2D_KinHel}~$e$), and disappear at a depth of about 500~km below the solar surface (panel $f$).

\section{Small-scale dynamo and links to low atmosphere layers}
From previous numerical studies it is known that important dynamical and energetic links between subsurface turbulent convective flows and the low atmosphere are established through small-scale vortex tubes. In the presence of magnetic field, the vortex tubes represent channels of energy exchange between the convective layers and the chromosphere and can result in the heating of chromospheric layers. This may be a source of small-scale spicule-like eruptions and Alfv\' en waves \citep{Kitiashvili2012,Kitiashvili2013a}.

In addition to these effects, our simulations of the local dynamo show that the anisotropy between the vertical and horizontal magnetic field components takes place only in the photosphere and above. The nature of this anisotropy has been the subject of  recent debate \citep[e.g][]{OrozcoSuarez2007,Lites2008,Danilovic2010,Ishikawa2010,Steiner2012,Stenflo2013}. The anisotropy changes with height, but there is no dependence on the `seed' field. The vertical distribution of the mean unsigned magnetic field components (Fig.~\ref{fig:Brms}$a$) shows a slow increase of their strength with depth that is a result of compression. In the convection zone, the distribution of the vertical and horizontal magnetic fields is statistically similar. This difference between the vertical and horizontal field strengths is small with a weak dominance of the horizontal fields below the solar surface. Indeed, the ratio between the transverse field, defined as $B_h=\sqrt{B_x^2+B_y^2}$, and the unsigned vertical field is only slightly above $\sqrt{2}$ (Fig.~\ref{fig:Brms}$b$), meaning that the strength of all the field components is similar. Near the surface layers the vertical magnetic flux becomes dominant. Above the photosphere the mean vertical unsigned flux slowly decreases, whereas the horizontal field increases and becomes dominant (Fig.~\ref{fig:Brms}).
 Note that our results are different from the previous simulations by \cite{Schussler2008} and \cite{Rempel2014} who found that the mean horizontal component of magnetic field is larger than the vertical component in the whole range, from the deep photosphere to the chromosphere. Our results indicate that the horizontal component is dominant only in the upper photosphere and low chromosphere, from about 300 km to 700 km.

This dominance of the vertical or horizontal fields in the different layers reflects the topological properties of the dynamo-generated magnetic fields, which are characterized by field lines organized in small-scale magnetic loops above the photosphere (Fig.~\ref{fig:B06_3D_m-lines}). Such a topological structure resembles the magnetic canopy suggested from observations  \citep{Giovanelli1980,Jones1982,Schrijver2002}. According to our simulation results, in the intergranular lanes magnetic field is mostly vertical and becomes horizontal above the photosphere. The height of these loops is greater when the magnetic field is stronger at the loop footpoints. Such a topology of the magnetic field lines was suggested both by observations \citep[e.g.][]{Lites2008} and simulations \citep[e.g.][]{Schussler2008,Steiner2008}. The horizontal magnetic fields are stronger at the granule edges and in the regions where strong turbulent motions are present. This tendency of the distribution of transverse magnetic fields was previously found in observations \citep{Lites2008}.  For a detailed comparison of the simulation results with the observational data it is necessary to perform spectro-polarimetric analyses for the simulated data, taking into account the instrumental characteristics. We plan this work for a future paper.

In addition, it is interesting to note that the closest opposite-polarity patches may not even be connected by magnetic field lines above the solar surface as simple magnetic loops, but instead interact through electric currents above and below the photosphere forming very complicated structures. Figure~\ref{fig:case3_3D_E-current} illustrates the topological structure of the electric current density streamlines above and below the photosphere (shown as a horizontal semi-transparent plane). Each streamline is tracked from a point in the region of positive-polarity (orange streamlines) and negative polarity (dark blue) patches. The topological structure of the electric currents above the photosphere is often characterized by spirals, arcs, and large swirls. Below the solar surface such a topology can represent highly turbulent flows, as in the case of the positive patch (orange streamlines, in Fig.~\ref{fig:case3_3D_E-current}), or as a very regular spiral structure, as in the case of the negative patch (blue lines). For instance, in Figure~\ref{fig:case3_3D_E-current} a current streamline originating in the positive polarity patch (orange) is strongly twisted around the negative patch.

\section{Discussion and conclusion}
In this paper, we addressed the problem of small-scale (local) dynamos that are probably responsible for a significant fraction of the background magnetic fields on the solar surface. To investigate this problem we used a 3D radiative MHD code and performed several simulation runs for different strengths ($10^{-6}$ to $10^{-2}$~G) and spatial distributions of the initial seed field  (Table~\ref{tab:cases}). After injection of the seed field, no magnetic flux was added to or removed from the computational domain.We find that, in all simulation cases, magnetic field generation is qualitatively different only during the first few minutes after initialization. This transient reflects differences in the initial seed fields. However, after about 4 hours of solar time the statistical distributions of the generated fields are similar. Therefore, in this paper we mostly focused on analysis of case $E$, in which the `seed' field was distributed as white noise with amplitude $\pm 10^{-6}$~G. Our simulations show that, due to turbulent dynamo action, the magnetic field can be locally magnified above the equipartition strength ($\sim 600$~G), reaching more than 2000~G in the photosphere.

Our simulation results show that magnetic field amplification is driven by converging flows into the intergranular lanes, shearing flows, and helical motions. All these mechanisms of magnetic field amplification are present in our numerical model and linked to each other. Thus, the local dynamo process represents a complicated interplay of multiple mechanisms that are difficult to separate from each other and that can contribute differently in individual magnetic field amplification events.

\begin{figure}
\begin{center}
\includegraphics[scale=1.2]{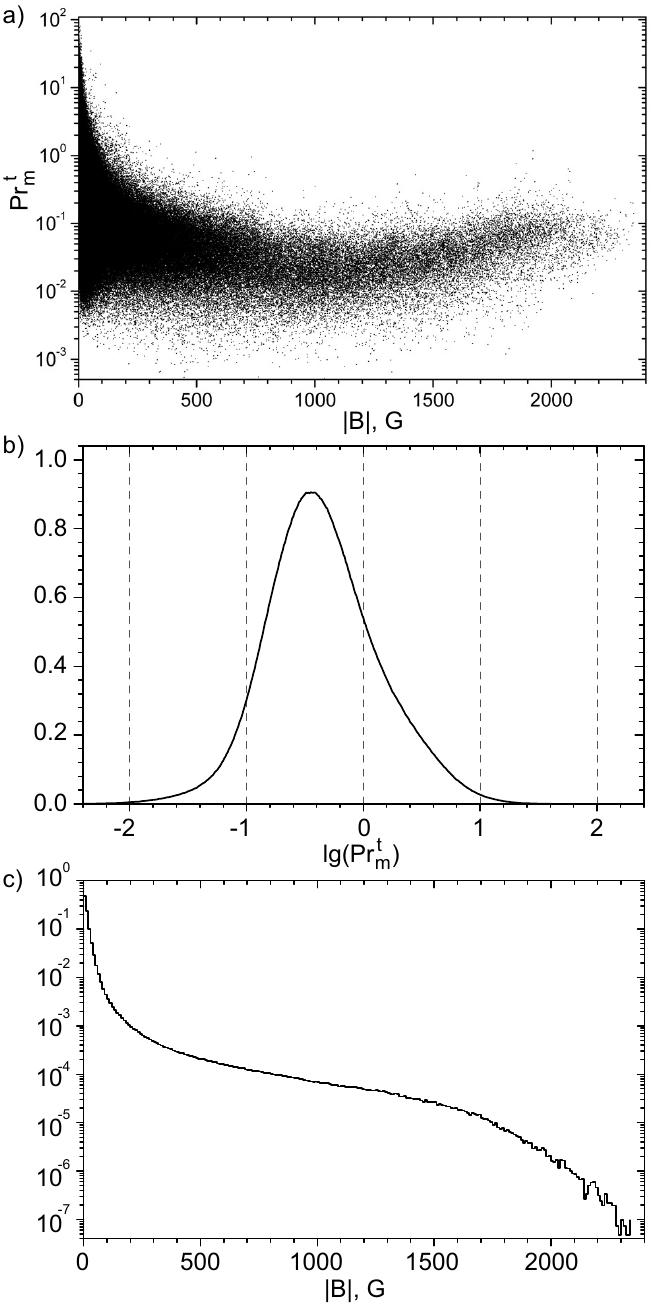}
\end{center}
\caption{$a$) Statistical distribution of the magnetic field strength ($|{\rm B}|$) vs. the turbulent magnetic Prandtl number ($Pr^t_m$). Probability distribution function (PDF) of $b$) Prandtl number (the statistical distribution parameters are: mean 0.143, variance 0.067, skewness 1.83 and kurtosis 2.011); and $c$) unsigned magnetic field.  \label{fig:hist1}}
\end{figure}

\begin{figure}
\begin{center}
\includegraphics[scale=1]{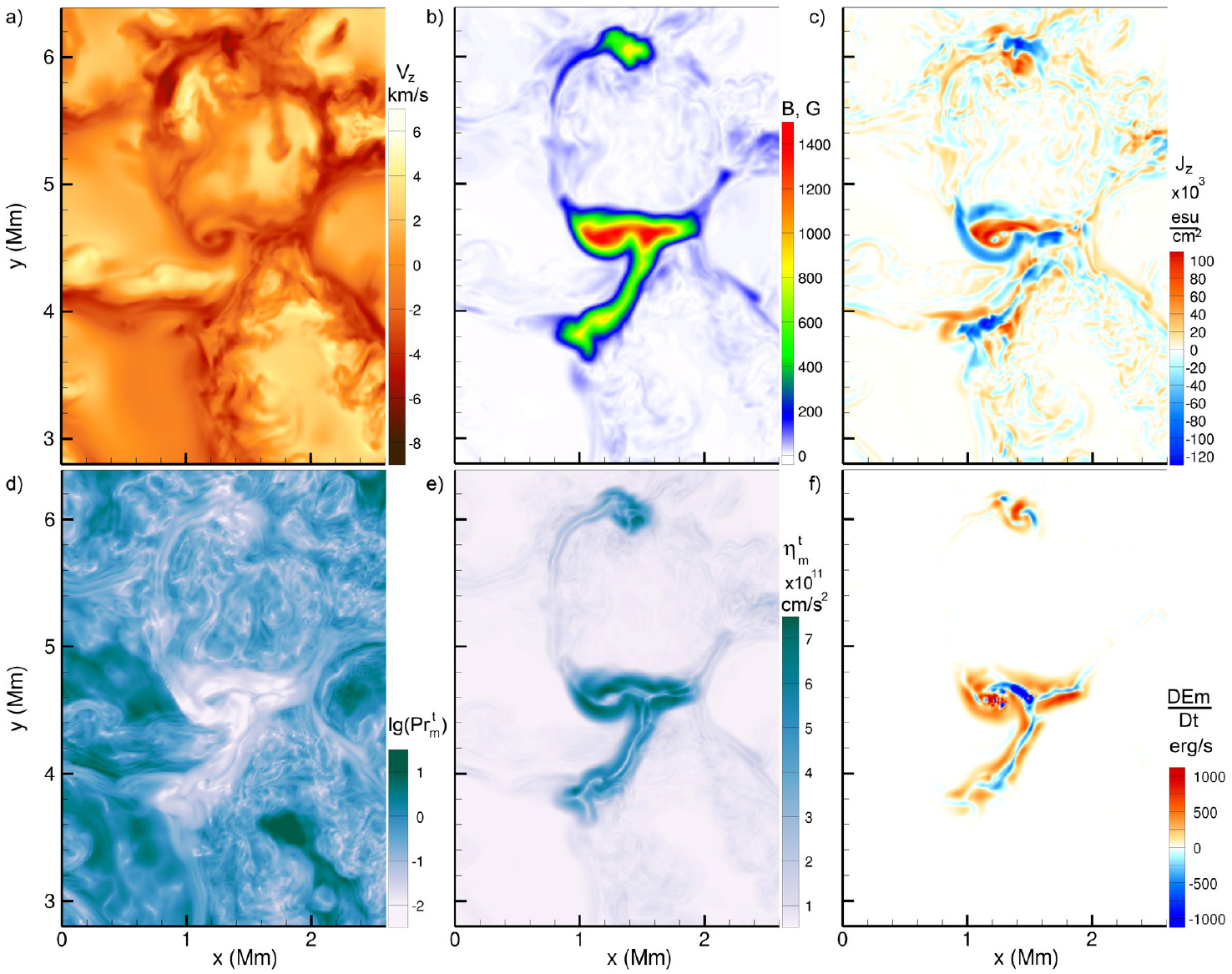}
\end{center}
\caption{Horizontal snapshot of a section of the computational domain in the photosphere for: $a$) vertical velocity; $b$) magnetic field strength; $c$) vertical component of the electric current density; $d$) logarithm of the turbulent magnetic Prandtl number; $e$) turbulent magnetic diffusivity; and $f$) time-derivative of the magnetic energy density. \label{fig:prm}}
\end{figure}

To model the local dynamo process we use the LES approach, which allows us to resolve the essential scales of the turbulent solar plasma and model subgrid scales. Unlike in many other dynamo simulations, our numerical setup does not restrict the magnetic turbulent Prandtl number $Pr_m^t$ to a constant value. For instance, Figure~\ref{fig:prm} illustrates a local distribution of various properties on the solar surface and shows a correlation between the areas of magnetic energy growth and small values of the magnetic Prandtl number. The statistical distribution of the turbulent magnetic Prandtl number shows that local variations can be of several orders of magnitude, but most of the distribution is in the range: $Pr_m^t \sim 10^{-1} - 1$ (Fig.~\ref{fig:hist1}$b$).

\begin{figure}
\begin{center}
\includegraphics[scale=0.95]{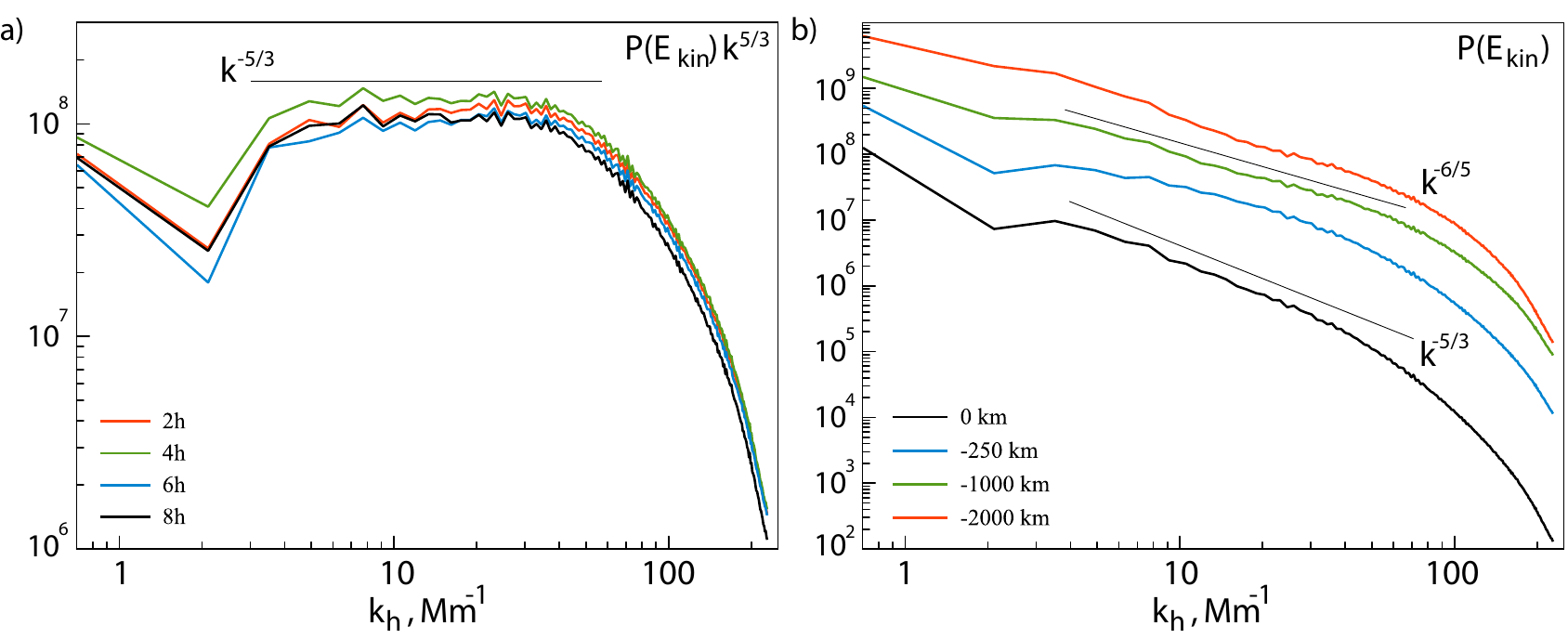}
\end{center}
\caption{Turbulent spectra of the kinetic energy density. Panel $a$: time-evolution of the kinetic energy spectra at the photosphere. Each curve corresponds to different moment of time: 2, 4, 6, and 8 hours after magnetic field initialization. The kinetic energy spectra are multiplied by a factor $k^{5/3}$. Panel $b$: turbulent spectra of the kinetic energy density at the saturation state for different depths. Each spectrum is averaged over 30~min. \label{fig:power1}}
\end{figure}

\begin{figure}
\begin{center}
\includegraphics[scale=0.9]{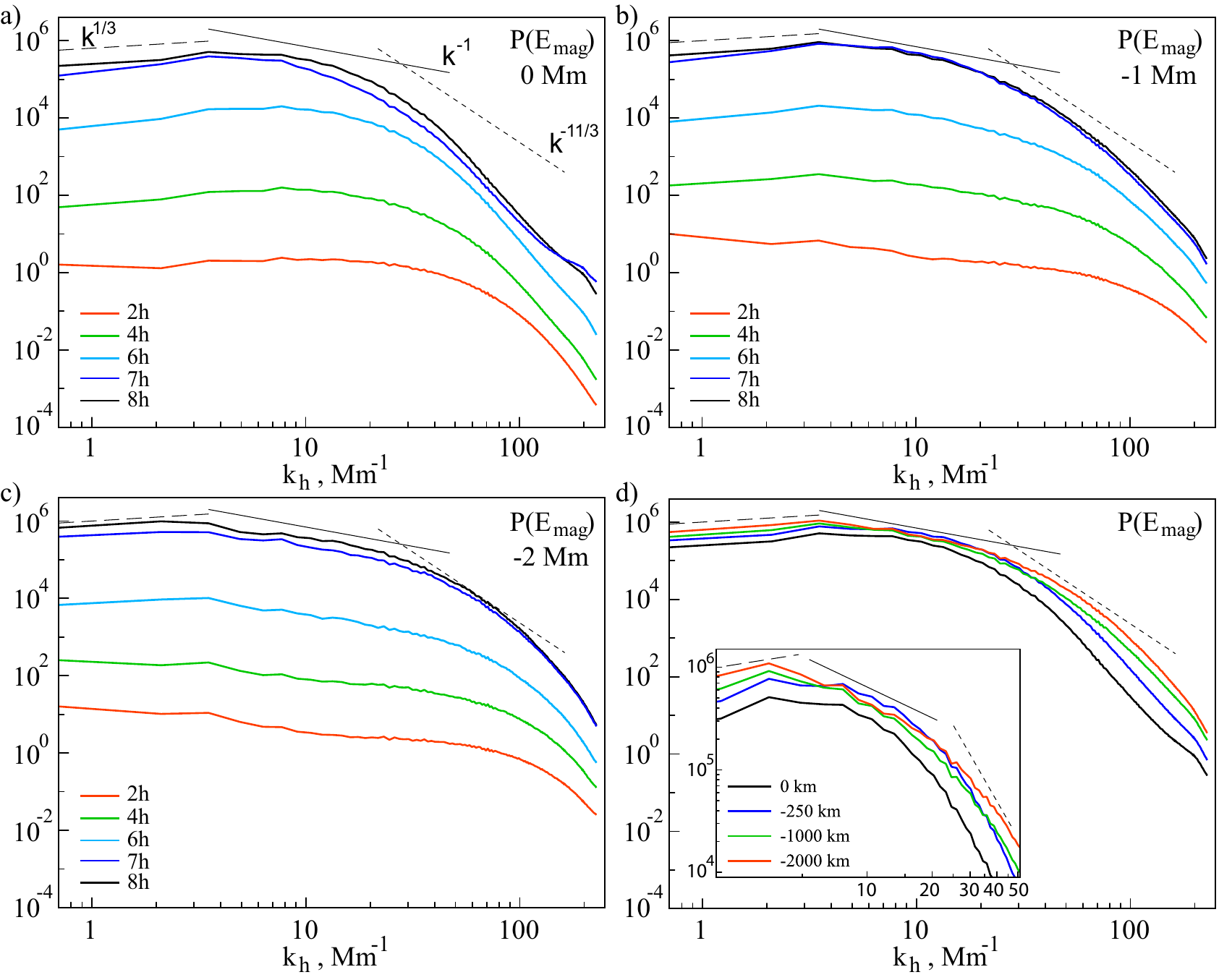}
\end{center}
\caption{Turbulent spectra for the magnetic energy density (case $E$). Panels $a - c$ show the evolution of magnetic energy density for different depths and times indicated in the figure. Panel $d$ shows a comparison of the magnetic energy spectra at the saturation state, for different depths. Each spectrum is averaged over 30~min. \label{fig:power}}
\end{figure}

The kinetic and magnetic energy-density spectra during the development of the dynamo process show an intense energy exchange between different scales and between the kinetic and magnetic energies (Figs~\ref{fig:power1} and~\ref{fig:power}). The energy redistribution causes changes in the slope of the power law, which in the developed regime reaches the Kolmogorov law ($k^{-5/3}$).

The development of dynamo action can be seen in the movie (in the supplementary material). Shortly after initialization of the `seed' field, magnetic patches appear in the photosphere. They are highly twisted and interact with each other. The magnetic energy in these patches increases with time. During the evolution of these magnetic elements strong helical downflows transport the magnetic field from the subsurface into deeper layers. Some of these patches diffuse; however, new magnetic patches are continuously formed. Our simulations show that the magnetic patches can have magnetic fields stronger than 2~kG (Fig.~\ref{fig:hist1}$c$).

\begin{figure}
\begin{center}
\includegraphics[scale=1.1]{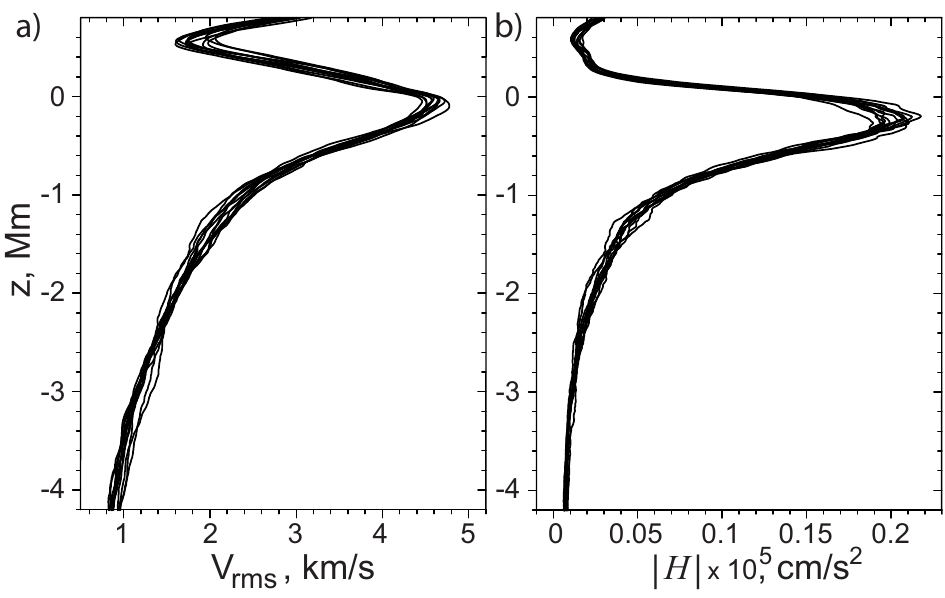}
\end{center}
\caption{Vertical profiles: $a$) rms velocity and $b$) kinetic helicity density plotted for 1 hour with  5~min cadence, 5~h after field initialization. \label{fig:time}}
\end{figure}

\begin{figure}
\begin{center}
\includegraphics[scale=1.]{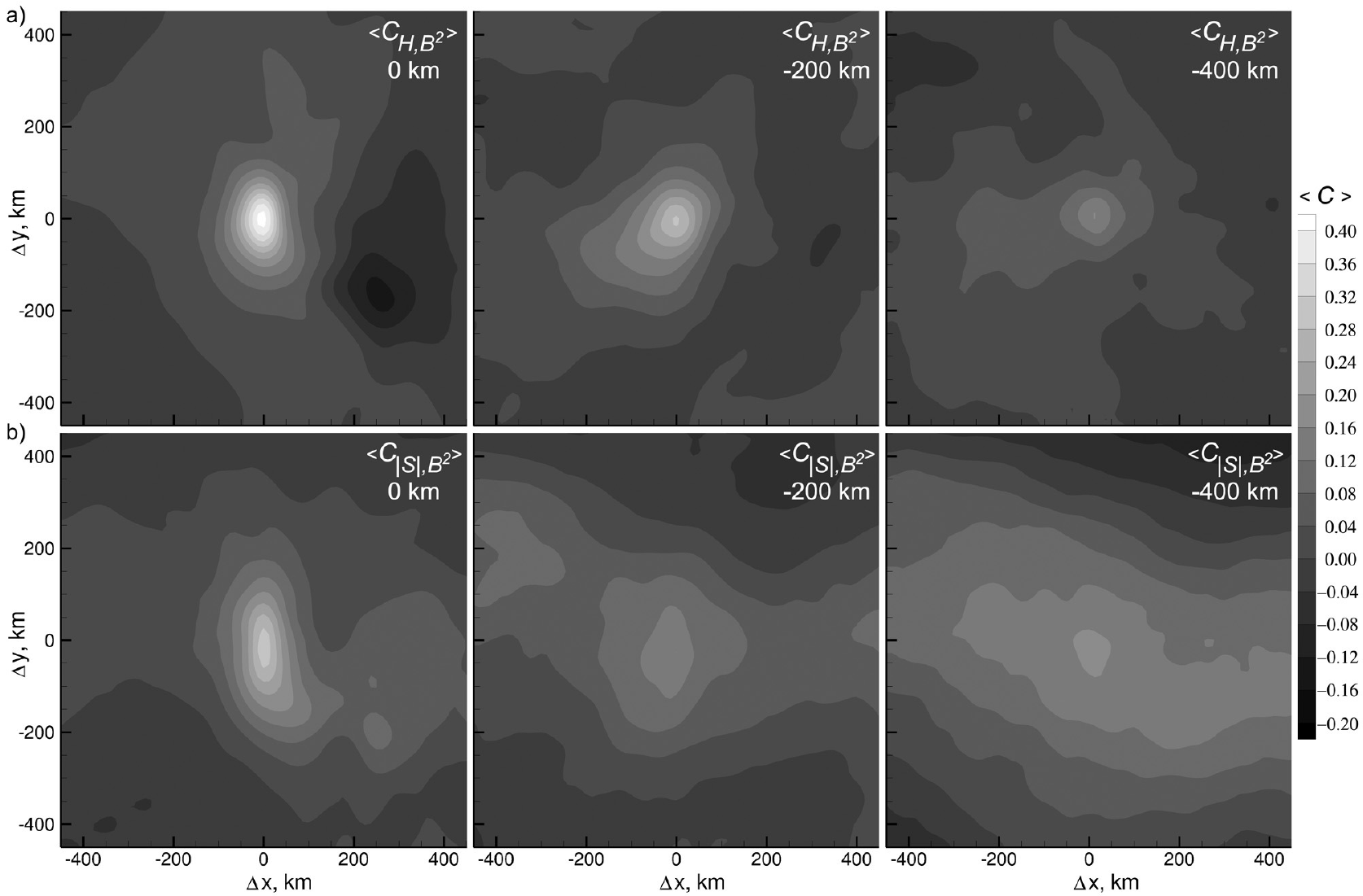}
\end{center}
\caption{Time-averaged distribution of cross-correlation coefficients of the kinetic helicity density, $H$ (top panels) and the  shear stress magnitude, $|S|$  (bottom panels) with the squared magnetic field, $B^2$, at three different depths: 0, $-200$, and $-400$~km. \label{fig:cross-corrHelU-B2}}
\end{figure}

Our analysis also shows that the local dynamo works most efficiently in the 1-Mm deep subsurface layer, where the turbulent flows are strongest (Figs~\ref{fig:time} and~\ref{fig:cross-corrHelU-B2}). We confirm the role of vortical motions of plasma in magnetic field amplification due to twisting and stretching \citep{Nordlund1992,Brandenburg1995,Brandenburg1996}. The magnetic patches generated due to shearing and vortical motions are transported into deeper layers by downdrafts. In the deeper layers the magnetic field can be further amplified by compression.

\begin{figure}
\begin{center}
\includegraphics[scale=0.73]{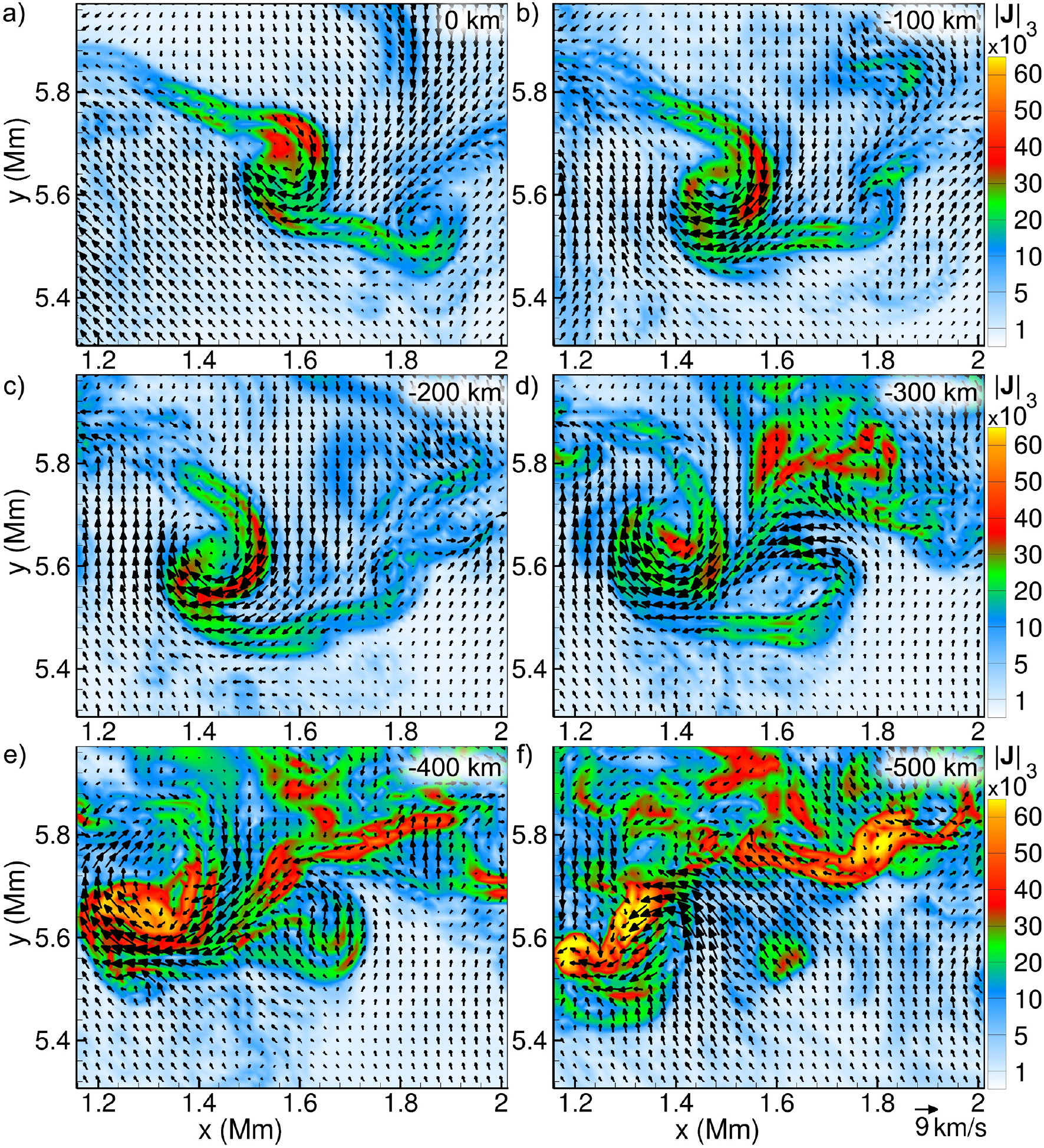}
\end{center}
\caption{The electric current density (color background) and the horizontal velocity field (arrows) at different depths,
from the photosphere (panel $a$) to 500~km below the photosphere (panel $f$) for the same moment of time as the first snapshot in Fig.~\ref{fig:case3_2D}.  \label{fig:case3_2D_E-current}}
\end{figure}

\begin{figure}
\begin{center}
\includegraphics[scale=0.73]{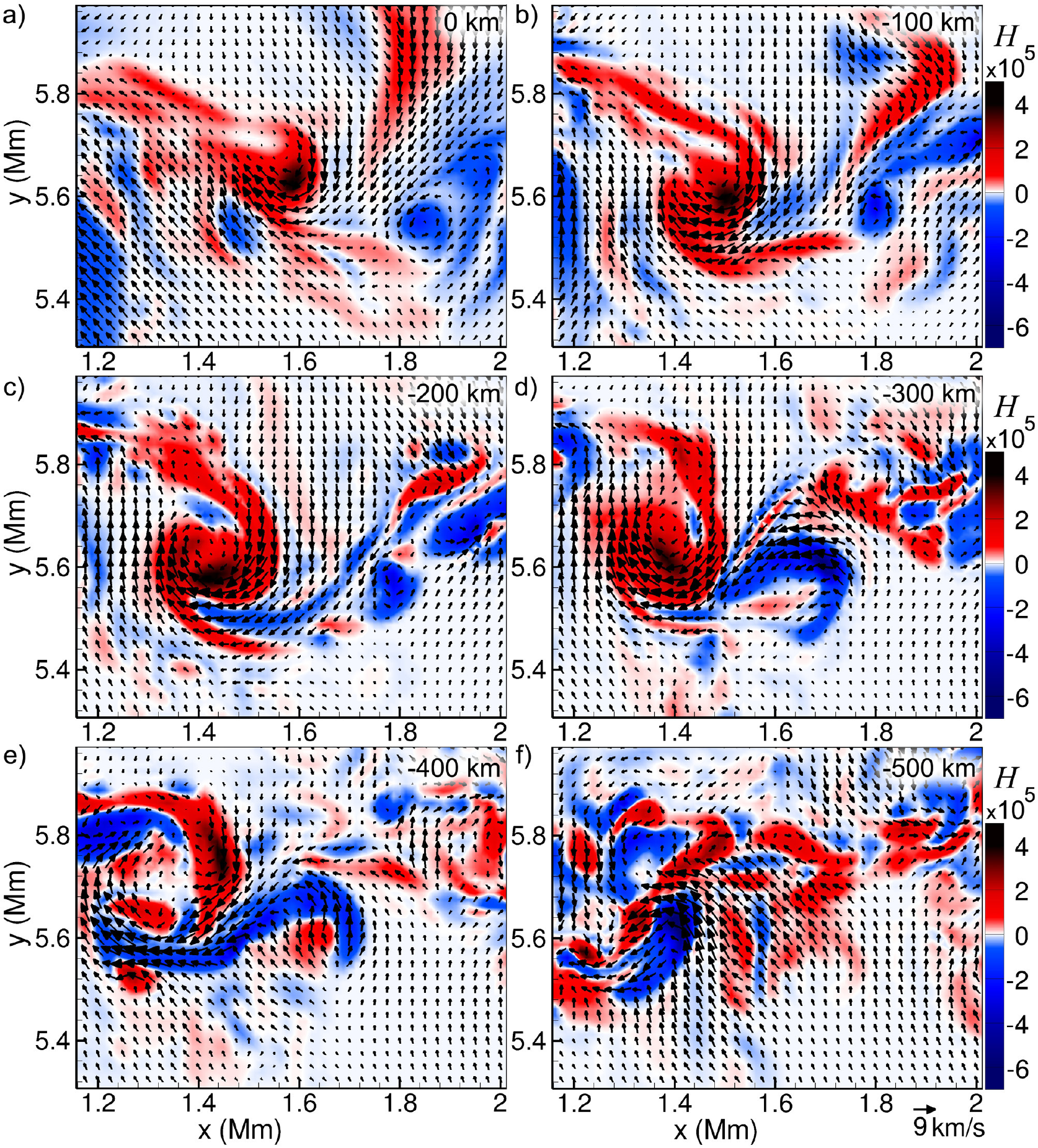}
\end{center}
\caption{The kinetic helicity density, $H$ (color background), and the horizontal velocity field (arrows)
at different depths from the photosphere (panel $a$) to 500~km below (panel $f$) for the same moment of time as in the first snapshot in Fig.~\ref{fig:case3_2D}. \label{fig:case3_2D_KinHel}}
\end{figure}

\begin{figure}
\begin{center}
\includegraphics[scale=1.]{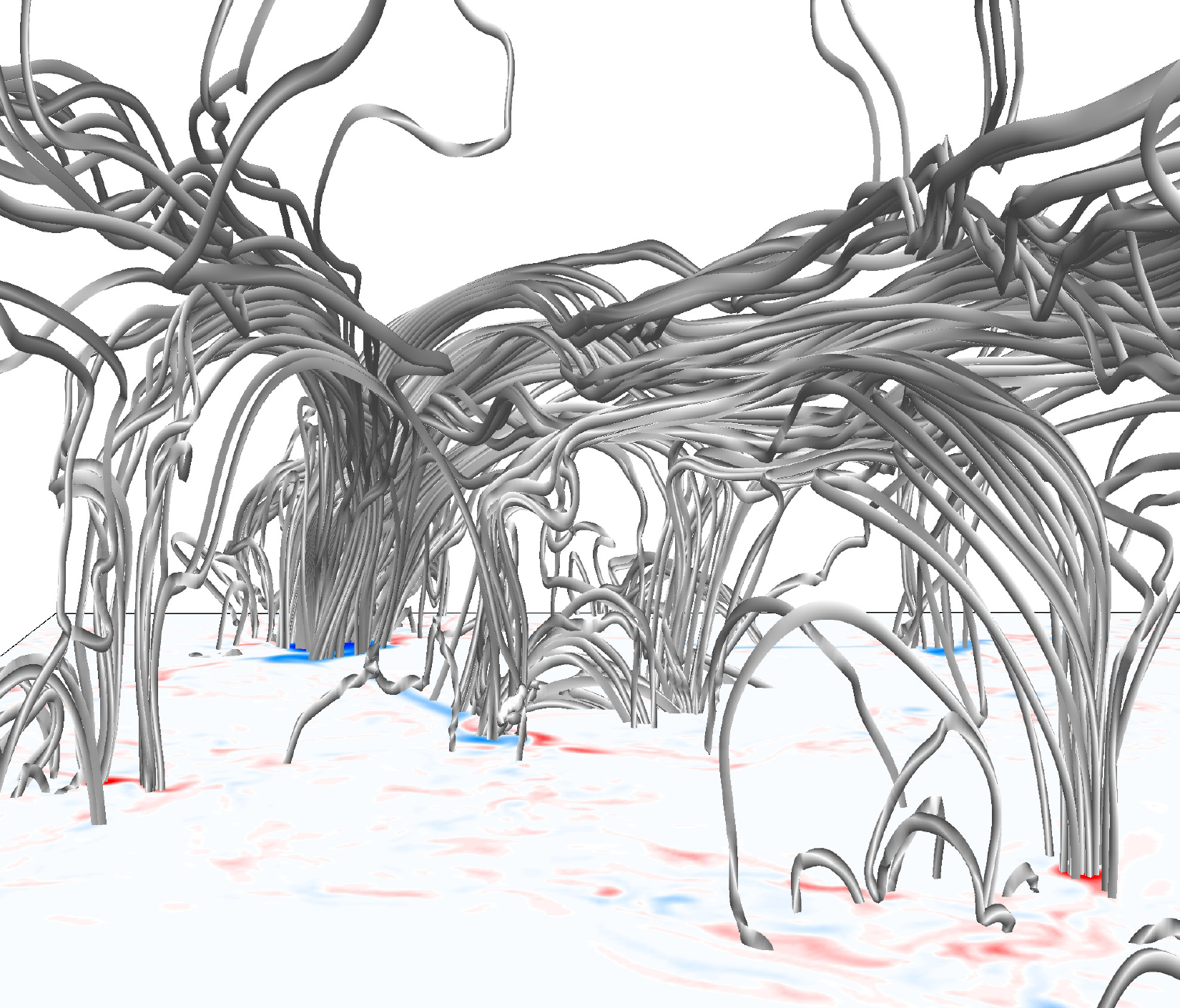}
\end{center}
\caption{Illustration of the topology of the magnetic field lines above the photosphere in the local dynamo simulations. The horizontal plane shows the distribution of the vertical magnetic field in the photosphere. Red color corresponds to positive polarity, blue color to negative polarity of the vertical magnetic field. The range of field strength is from $-800$~G to 300~G. \label{fig:B06_3D_m-lines}}
\end{figure}

\begin{figure}
\begin{center}
\includegraphics[scale=0.75]{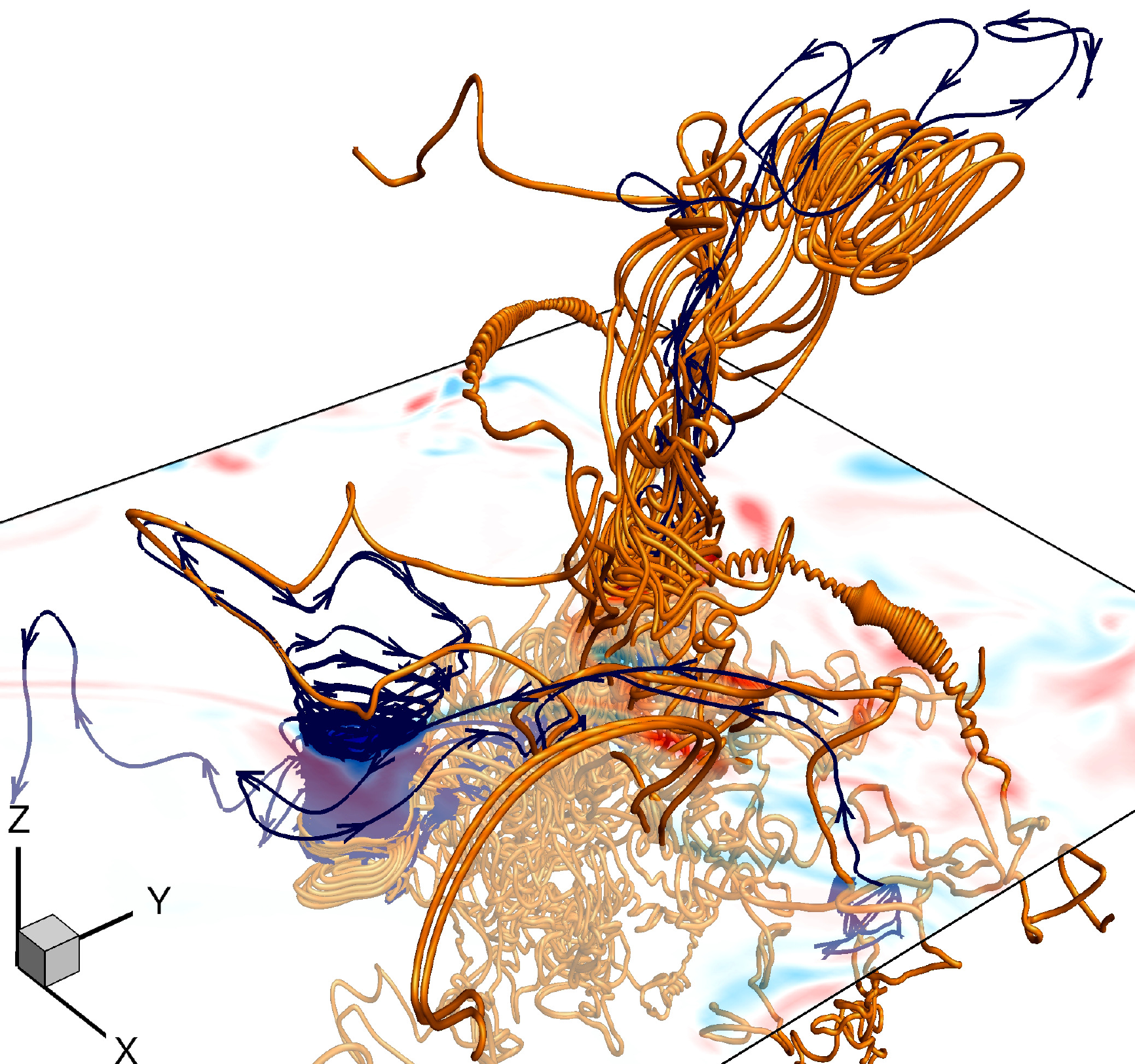}
\end{center}
\caption{Example of the topological structure of the electric current density below and above the photosphere.
Streamlines correspond to electric currents originating from the positive (orange) and negative (blue) polarity patches. The semi-transparent horizontal plane shows the vertical magnetic field distribution in the photosphere, where blue color indicates negative polarity and red color positive polarity. The snapshot corresponds to the rightmost frames of Figure~\ref{fig:case3_2D}.  \label{fig:case3_3D_E-current}}
\end{figure}

We presented two characteristic examples in detailed studies: a bipolar magnetic structure generated by the interaction of vertical and horizontal vortex tubes, and magnetic field generation by a combined action of a vertical vortex tube and shearing flows. These examples illustrate three important properties of the process of magnetic field amplification. The first is a strong coupling of various magnetic field amplification mechanisms, which leads to increased local magnetic energy. The second key property of local dynamo action is the multi-scale nature of this process, which evolves in turbulent flows from the smallest resolved scales to the granular scales (Figs~\ref{fig:case3_2D_E-current} and~\ref{fig:case3_2D_KinHel}). The third property is the complex topological and dynamical structure of the process, illustrated in Figure~\ref{fig:case3_3D_E-current}, and which is reflected in the interaction of individual magnetic patches and surrounding magnetic fields and in self-organization into a magnetic network. In this paper we described just some of these properties of magnetic field evolution; further investigation is required. The primary topology of the dynamo-generated magnetic field is represented by compact magnetic loops appearing as bipolar structures in the intergranular lanes (Fig.~\ref{fig:B06_3D_m-lines}); these loops reach higher levels for stronger magnetic field concentrations at the footpoints of the loops.

\begin{figure}
\begin{center}
\includegraphics[scale=1]{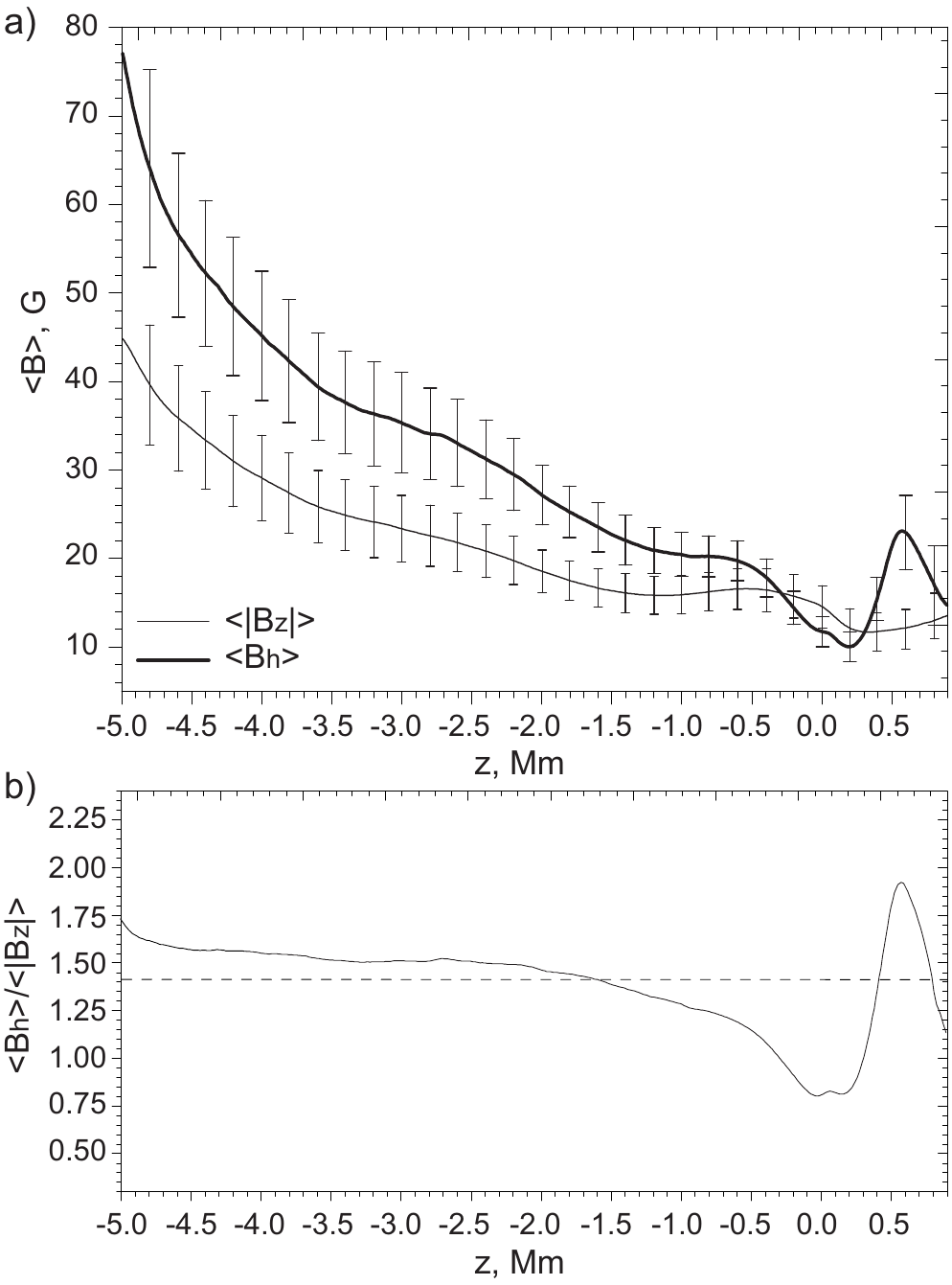}
\end{center}
\caption{$a$) Mean vertical profiles of the unsigned magnetic field components (averaged over 1 hour): vertical component (red curve) and transverse component (blue). Error bars show the standard deviation. $b$) Ratio of the mean vertical and transverse components of magnetic field as function of depth below the photosphere. Dashed line in panel $b$) corresponds to value $\sqrt 2$. \label{fig:Brms}}
\end{figure}

Formation of magnetic loops in the solar atmosphere reflects the observed height-dependent anisotropy of vertical and transverse small-scale magnetic fields (Fig.~\ref{fig:Brms}). We found an equipartition of the field component in the layers deeper than 2-Mm below the solar surface, but above the solar surface the transverse magnetic fields are dominant because of compact loop topology. This variation of the vertical and horizontal field anisotropy can explain discrepancies among different observations \citep[e.g][]{OrozcoSuarez2007,Lites2008,Danilovic2010,Ishikawa2010,Stenflo2013} and support previous numerical analysis \citep{Schussler2008,Steiner2008}.

Investigation of the small-scale dynamo action is a key element for understanding dynamical and eruptive processes on the Sun and forms one of the building blocks of the global dynamics. In this paper we suggest a new view of the problem through analysis of realistic-type radiative MHD simulations of the solar subsurface and chromosphere. We plan to continue analysis of these data, make detailed comparisons with observations, and clarify the physics of the small-scale dynamo action and organization of the Sun's magnetic network.

{\it\bf  Acknowledgements.} The simulation results were obtained on NASA's Pleiades supercomputer at NASA Ames Research Center. This work was partially supported by NASA grants NNX10AC55G and NNH11ZDA001N-LWSCSW, and Oak Ridge Associated Universities.

%\newpage

Y

\end{document}